\documentclass[10pt, twocolumn, comsoc]{IEEEtran}

\def\argmax{\mathop{\rm arg\,max}}
\def\argmin{\mathop{\rm arg\,min}}
\usepackage{epsfig}
\usepackage[noadjust]{cite}
\usepackage{mcite}
\usepackage{gensymb}
\usepackage{algpseudocode}
\usepackage{amsfonts,helvet}
\usepackage{fancyhdr}
\usepackage{threeparttable}
\usepackage{epsf,epsfig}
\usepackage{amsthm}
\usepackage{amsmath}
\usepackage{siunitx}
\usepackage{amssymb}
\usepackage{dsfont}
\usepackage{subfigure}
\usepackage{color}
\usepackage{multicol,amsmath,lipsum}
\usepackage{graphicx}
\usepackage{epstopdf}
\usepackage[margin=2cm]{geometry}
\usepackage[labelformat=simple]{subcaption}
\usepackage{stfloats}
\usepackage[linesnumbered,ruled]{algorithm2e}
\usepackage{enumerate}
\usepackage{cancel}
\usepackage{bm}
\usepackage{subfigure}
\usepackage{epstopdf}
\usepackage{comment}

\usepackage{siunitx}
\usepackage{lettrine}

\newtheorem{theorem}{Theorem}

\newtheorem{corollary}{Corollary}

\newtheorem{lemma}{Lemma}

\newtheorem{proposition}{Proposition}
\newtheorem{remark}{Remark}

\usepackage{amsthm}

\SetKwInput{KwInput}{Initialize}
\SetKwInput{KwOutput}{Output}
\SetKwProg{Fn}{Stage 1}{}{}
\SetKwProg{Fn}{Stage 2}{}{}

\usepackage{eucal}

\begin{document}

\title{Splitting Messages in the Dark-- \\ Rate-Splitting Multiple Access for FDD Massive MIMO without CSI Feedback}

\author{Namhyun~Kim, Ian P. Roberts, and Jeonghun~Park   

\thanks{

N. Kim and J. Park are with the School of Electrical and Electronic Engineering, Yonsei University, Seoul 03722, South Korea. (e-mail: {\text{namhyun@yonsei.ac.kr, jhpark@yonsei.ac.kr}}).
I. P. Roberts is with the Department of Electrical and Computer Engineering, University of California,
Los Angeles, CA 90095 USA. (e-mail: {\text{ianroberts@ucla.edu}}).
}}

\maketitle \setcounter{page}{1} 

\begin{abstract} 
    A critical hindrance in realizing frequency division duplex (FDD) massive multi-input multi-output (MIMO) systems is the overhead associated with the downlink (DL) channel state information at the transmitter (CSIT) acquisition. To address this, we propose a novel framework that eliminates the need for CSI feedback, while achieving robust sum spectral efficiency (SE). Specifically, by leveraging partial frequency invariance of channel parameters, we reconstruct the DL CSIT using uplink (UL) pilots with the 2D-Newtonized orthogonal matching pursuit (2D-NOMP) algorithm. \textcolor{black}{Due to discrepancies between the two disjoint bands, however, perfect DL CSIT acquisition is infeasible; resulting in multi-user interference (MUI).} To account for this, we reformulate the sum SE maximization problem using the reconstructed channel and its error covariance matrix (ECM). Then, we propose an ECM estimation method based on the observed Fisher information matrix and introduce a precoder optimization technique with rate-splitting multiple access (RSMA). 
    Our simulation results verify the validity of the proposed framework in the practical FDD massive MIMO scenarios, highlighting the essential role of ECM estimation in mitigating MUI to attain RSMA gains.

\end{abstract} 



\section{Introduction}
Massive multiple-input multiple-output (MIMO) systems are key to the future of wireless communications, effectively increasing spectral efficiency (SE) and reducing costs \cite{Overview_MIMO, 6G-era}. To fully realize these benefits, accurate channel state information at the transmitter (CSIT) is critical. In this context, time division duplexing (TDD)-based channel estimation has been preferred for its efficiency from channel reciprocity \cite{FDD-measurement}.

Nonetheless, there have been persistent research efforts aimed at realizing the advantages of massive MIMO for frequency division duplexing (FDD) systems as well \cite{T.Choi:2021, Rottenberg:2020}. 
The enduring importance of these studies is usually underscored by the following considerations.
i) The majority of the bandwidth is allocated to the FDD mode in sub-6 GHz, which still dominates the cellular market today \cite{junil:jstsp:14}.
ii) FDD typically offers superior coverage performance compared to TDD \cite{FDD-measurement}, allows simultaneous uplink (UL) and downlink (DL) transmissions, catering to delay-sensitive applications \cite{FDD-delay}.
Despite these benefits, FDD massive MIMO systems face significant challenges as the number of base station (BS) antennas grows, primarily due to the prohibitive complexity of CSIT acquisition from the lack of channel reciprocity \cite{JOMP, Overview_MIMO}. \textcolor{black}{In this paper, we take a step toward addressing this problem while maximizing the sum SE performance.}



\subsection{Prior Work}
We categorize the prior work on realizing FDD massive MIMO systems into two primary approaches: those aimed at reducing CSIT overhead and those aimed at mitigating multi-user interference (MUI).

\subsubsection{CSIT overhead reduction approaches}
One of the main principles of these studies is that the channel characteristics in FDD MIMO can be compressed in certain domains, which is helpful to reduce the overhead \cite{CS-love}. 
For instance, a distributed DL CSIT compression method was developed by \cite{JOMP}, which harnessed the shared sparsity structures from local scatterers.
In \cite{CS-OFDM}, a super-resolution compressed sensing (CS) method was proposed in the presence of the beam squint effect in mmWave wideband MIMO-OFDM systems. 
Recently, methods not relying on a specific compressible basis but rather on data-driven approaches have also been studied. 
For example, in \cite{CS-deep}, CS-ReNet was devised, which jointly exploits deep learning techniques and CS-based compression.
For example, \cite{DLCE:JSAC:21} proposed a multiple-measurement-vectors learned approximate message passing (MMV-LAMP) network to enable efficient channel reconstruction of the spatial-frequency channel matrix, using channels’ structured sparsity.
In \cite{Deeplearning-Ahmed}, it was proposed to use deep learning to map channels from one set of antennas and frequency band to channels of another set, significantly reducing the training overhead. 
An extensive overview of deep learning based CSIT acquisition was provided in \cite{guo:tcom:22}.

In a different line of research, \cite{Dina} presented an intriguing idea: the DL channel in FDD can be fully reconstructed by UL training without CSI feedback. 
Specifically, DL channel parameters, such as the angle of departure (AoD), delay, and path gains, can be obtained from the UL channel, assuming that the frequency difference of UL and DL is small enough relative to the carrier frequency. Extensions of this concept were discussed in \cite{T.Choi:2021, Rottenberg:2020, Zhong:2020, Deokhwan}, with \cite{Han:2019} supported by measurement campaigns. {\textcolor{black}{
In \cite{An:ml:23}, channel quality indicator (CQI) feedback was omitted by exploiting a machine learning (ML) based technique for reducing delay.}}
However, the accuracy of the reconstructed DL channel is sometimes reported to be limited due to discrepancies between the UL and DL channels, as shown in \cite{T.Choi:2021, Zhong:2020}. It is important to note that this imperfection can significantly undermine the SE performance in DL due to increased MUI. 
\subsubsection{MUI mitigation approaches}
In the presence of imperfect CSIT, 
MUI mitigation strategies have been explored as another approach to unleash the gains of FDD massive MIMO systems. In \cite{MAT}, an interference alignment scheme was developed to manage MUI caused from delayed CSIT. 
In \cite{JSDM}, it was shown that when users are sufficiently spatially separated, their channel covariance matrices have non-overlapping subspaces, leading to reduction of MUI. 


Recently, an unorthodox type of multiple access, referred as rate-splitting multiple access (RSMA), has also been actively explored to handle the MUI problem \cite{RSMA-ten-promising}. 
In RSMA, users partially decode MUI through the introduction of a common message and successive interference cancellation (SIC), leading to interference decoding gains. 
Building on this, \cite{WMMSE-SAA} presented a linear precoder design based on the weighted minimum mean square error (WMMSE) method, using the sample average approximation (SAA) technique to account for imperfect CSIT in RSMA. 
Beyond this, \cite{Park:2023} proposed a generalized power iteration (GPI)-based precoder optimization, assuming that the channel covariance and DL channel error covariance are given in TDD MIMO. In addition, a bilinear precoding method in FDD was introduced in \cite{Amor:2023}, considering that the knowledge of channel covariance is provided.
\textcolor{black}{
A fractional programming (FP) based RSMA precoding method was developed in \cite{mao:twc:24}, wherein the optimal beamforming structure of RSMA is incorporated. }

Notice that the two approaches—CSIT overhead reduction and MUI mitigation—are closely related; yet most previous studies have largely existed in isolation, focusing on one approach while overlooking the other.
{\color{black}{For instance, efficient linear precoding methods for RSMA were explored in several prior work by considering various CSIT acquisition models; such as FDD MIMO with limited feedback \cite{mao:twc:24} or FDD MIMO with analog feedback \cite{Amor:2023}. However, no prior work has explored the case where CSI feedback is entirely omitted in FDD MIMO. A key challenge in this case is that, since the DL CSIT cannot be directly obtained, it must be reconstructed from side information, such as UL reference signals \cite{Dina, Han:2019}. 
Nonetheless, the impact of the DL CSIT reconstruction algorithm on RSMA performance is not clear \cite{Dina, Han:2019}, thus it is not yet known how to incorporate such DL CSIT reconstruction into RSMA precoder design. 
For this reason, jointly tackling the DL channel reconstruction and the RSMA precoder design remains an open problem.
}}

\subsection{Contributions}
{\color{black}{In this paper, we put forth a comprehensive approach to reduce the CSIT acquisition overhead and also efficiently mitigate MUI. 
Specifically, we consider an FDD massive MIMO system without direct CSI feedback, and instead estimate the DL channel based on UL pilots. 
To accomplish this, we employ the 2D-Newtonized orthogonal matching pursuit (NOMP) algorithm that extracts key parameters of UL channels, such as angle of arrival (AoA), delay, and path gain \cite{Han:2019}. 
With this, we rebuild the DL channel by leveraging partial frequency invariance between UL and DL channels \cite{Zhong:2020, T.Choi:2021, Han:2019}. 
However, due to the inherent discrepancies between the UL and DL channels, CSIT errors are unavoidable, which results in MUI that undermines the SE gains. To resolve this, we employ RSMA. }}

{\color{black}{Nonetheless, realizing the full potential of RSMA is not straightforward. This is mainly because it is difficult to capture the impact of imperfect CSIT reconstruction on RSMA performance, making it challenging to efficiently design the precoder that accounts for the CSIT errors. 
To resolve this, we characterize the achievable SE as a function of the precoding vectors, the reconstructed CSIT, and the CSIT error covariance matrix (ECM). 
In our characterization, we find that the ECM plays a crucial role in designing precoders that effectively mitigate the MUI. 
Since the ECM cannot be obtained in a conventional way due to non-linearity of 2D-NOMP, we develop a novel ECM estimation method. Our rationale is built on the Cramér-Rao lower bound (CRLB) of DL channel reconstruction, where its trace provides a lower bound for MSE. However, computing the CRLB requires true channel parameters, which is not feasible in practice. 
To address this, we use the observed Fisher information matrix (O-FIM) \cite{efron1978assessing,observedFisher,Observed-Relative}, which requires only the observed UL reference signals. This enables us to compute the CRLB without the ground truth value or CSI error distribution, thereby improving the practicability of our method.}}

{\color{black}{
Subsequently, harnessing the estimated ECM, we propose a precoder optimization method to solve the sum SE maximization problem. Specifically, we derive the Karush-Kuhn-Tucker (KKT) condition for maximizing the achievable sum SE. This can then be interpreted as an eigenvector-dependent nonlinear eigenvalue problem. Accordingly, by finding the leading eigenvector of the derived KKT condition, we reach a locally optimal solution that maximizes the sum SE. 
}}

{\color{black}{
Through numerical simulations, we demonstrate that the proposed method offers significant improvement in terms of sum SE compared to existing methods. For instance, we observe up to a 22.7\% increase in sum SE performance compared to other baseline spatial division multiple access (SDMA) methods when no CSI feedback is used.
Crucially, we reveal that these gains primarily stem from our ECM estimation. 
By accurately estimating the ECM and incorporating this into the precoder optimization, we properly account for the CSIT errors and the resultant MUI. 
Specifically, when incorporating the estimated ECM into the precoder design, we observe up to a 21.1\% improvement in sum SE performance compared to a case that does not rely on the estimated ECM. 
In addition, we also examine a case where the number of DL channel paths is reduced compared to the UL. We observe that the proposed method still provides robust performance with our appropriate ECM estimation.
Further, we present that our framework unlocks new possibilities for enabling low-latency FDD massive MIMO transmission. Specifically, we analyze the latency overhead associated with the typical 5G-NR process of CSI feedback. Based on this, we compare the achievable sum SE between the perfect CSIT case (with feedback) and the no CSI feedback case, offering insights into the trade-off between latency overhead and achievable sum SE. 
This trade-off highlights how our framework provides an effective alternative for conventional FDD massive MIMO transmission, significantly reducing latency with only a marginal sacrifice in SE performance. 
}}

\textcolor{black}{\textit{Notation}:
The superscripts $(\cdot)^{\sf T}$, $(\cdot)^{\sf H}$, $(\cdot)^{-1}$, $(\cdot)^{\dagger}$ denote the transpose, Hermitian, matrix inversion, and Moore-Penrose pseudo-inverse respectively. ${\bf{I}}_N$ is the identity matrix of size $N \times N$, Assuming that ${\bf{A}}_1, ..., {\bf{A}}_N \in \mathbb{C}^{K \times K}$, ${\bf{A}} = {\rm diag}\left({\bf{A}}_1, ...,{\bf{A}}_n,..., {\bf{A}}_N \right)$ is a block-diagonal matrix concatenating ${\bf{A}}_1, ..., {\bf{A}}_N$. We use $\lfloor\cdot\rfloor, \lceil\cdot\rceil$ to denote rounding a decimal number to its nearest lower and higher integers, respectively. We define $[K]$ as the set of natural numbers less than or equal to $K$.
$\mathbf{A}\circ\mathbf{B}$ denotes the Hadamard product of the two matrices, $\mathbf{A}$ and $\mathbf{B}$.}

\section{System Model} \label{sec:model}
We consider the multi-user FDD massive MIMO system, where the BS is equipped with $N$ antennas and serves $K$ single-antenna users. We also assume that the channel remains time-invariant from the transmission of the UL pilots to the DL precoding \cite{Han:2019, Rottenberg:2020}. 

\subsection{Channel Model}

\textcolor{black}{With orthogonal frequency division multiplexing (OFDM) employed, we denote the number of sub-carriers used for pilots as \( M \) and the spacing between these sub-carriers as \( \Delta f \) which implicitly defines the bandwidth of each resource block. That is, these \( M \) pilots are uniformly spaced, and the total bandwidth they span is \(B = M\Delta f \).   
Following the multipath channel model in \cite{Han:2019, Dina}, we represent the UL channel on the $m$-th sub-carrier of user $k \in [K]$ as \
\begin{align}
    \boldsymbol{\mathbf{h}}^{\mathrm{ul}}_{k}[m] = \sum \limits^{L_k^{\mathrm{ul}}}_{\ell=1}\mathrm{\alpha}^{\mathrm{ul}}_{k, \ell}\boldsymbol{\mathrm{a}}\left(\theta^{\mathrm{ul}}_{k, \ell}; {\lambda^{\mathrm{ul}}_m}\right)e^{-j2\mathrm{\pi}m\Delta f \tau^{\mathrm{ul}}_{k, \ell}} \in \mathbb{C}^{N \times 1}, \label{eq:ul channel}
\end{align}}
\textcolor{black}{
\noindent where $\mathrm{\alpha}^{\mathrm{ul}}_{k, \ell}$ is the complex path gain of the $\ell$-th path, and ${\mathbf{a}}(\theta^{\mathrm{ul}}_{k, \ell}; {\lambda^{\mathrm{ul}}_m)}$ is the ULA array response defined sub-carrier specifically as}
\textcolor{black}{\begin{align} \label{eq:ul_model}
{\mathbf{a}}(\theta^{\mathrm{ul}}_{k, \ell};{\lambda^{\mathrm{ul}}_m)} = \left [1, e^{-j2\pi \frac {d^{\mathrm{ul}}}{\lambda^{\mathrm{ul}}_m} \sin \theta^{\mathrm{ul}}_{k, \ell}}, \ldots , e^{-j2\pi {(N-1)} \frac {d^{\mathrm{ul}}}{\lambda_m^{\mathrm{ul}}} \sin \theta_{k, \ell}^{\mathrm{ul}}}\right]^{\sf{T}},\end{align}
where we denote the AoA as $\theta^{\mathrm{ul}}_{k, \ell}$, the wavelength of the $m$-th UL sub-carrier as $\lambda_m^{\mathrm{ul}}$, and the antenna spacing as $d^{\mathrm{ul}}$ with $d^{\mathrm{ul}} = \lambda_{\mathrm{c}}^{\mathrm{ul}}/2 $, where $\lambda_{\mathrm{c}}^{\mathrm{ul}}$ is the wavelength of carrier frequency of UL.}
Further, we denote the propagation delay as $\tau^{\mathrm{ul}}_{k, \ell}$ with $0<\tau^{\mathrm{ul}}_{k, \ell}<1/\Delta f$ and the number of paths as $L_k^{\mathrm{ul}}$. 
We also assume that the baseband UL signal occupies the interval $[-B/2, B/2]$\footnote{This follows the convention where the carrier frequency of the UL band is set to zero \cite{Rottenberg:2020}. Accordingly, we can represent the DL channel as a function of the frequency difference between the DL and UL.}. Correspondingly, index $m$ satisfies $\lfloor \frac{-M}{2} \rfloor \leq m \leq \lceil \frac{M}{2} \rceil-1, m \in \mathbb{Z}$.


To model the distribution of the complex path gain $\mathrm{\alpha}^{\mathrm{ul}}_{k, \ell}$, we use the Rician distribution, i.e. $\mathrm{\alpha}^{\mathrm{ul}}_{k, \ell} \sim \mathcal{CN}\left(\sqrt{\frac{\kappa_k\sigma^2_{\mathrm{path}, k}}{\kappa_k+1}}, \frac{\sigma^2_{\mathrm{path}, k}}{\kappa_k+1}\right), \forall \ell$.
The parameter $\kappa_k$ denotes the Rician factor of user $k$ and represents the relative strength of the line-of-sight (LoS) component compared to the non-line-of-sight (NLoS) component. The parameter $\sigma^2_{\mathrm{path}, k}$ denotes the average channel power of each path, i.e. $\mathbb{E}[|\mathrm{\alpha}^{\mathrm{ul}}_{k, \ell}|^2] = \sigma^2_{\mathrm{path}, k}, \forall \ell,$ where we define $\sigma^2_{\mathrm{path}, k} = 1/(NL^{\mathrm{ul}}_k)$ to normalize the channel in \eqref{eq:ul channel} to unit variance.
We note that $\alpha_{k, \ell}, \theta_{k, \ell}, \tau_{k, \ell}$ are consistent across the sub-carriers within operating bandwidth, under the assumption that the bandwidth is much smaller than the carrier frequency \cite{Han:2019, Rottenberg:2020}.

{\color{black}{
Notice that in \eqref{eq:ul channel}, we describe the UL channel vector ($\mathbf{h}_k^{\mathrm{ul}}[m]\in\mathbb{C}^{N\times 1}$) with respect to the OFDM sub-carrier index $m$. This signal representation in the frequency domain is crucial for the joint estimation of angles ($\theta_{k, \ell}^{\mathrm{ul}}$) and delays ($\tau_{k, \ell}^{\mathrm{ul}}$) during the UL training phase (as explained in Section III) as demonstrated in \cite{Dina}. 
{In contrast, in other parts—such as defining performance metric and designing precoders—the sub-carrier notation becomes irrelevant. Once reconstructed DL channel is obtained during UL training, the performance metric, i.e., SE, depends solely on the composite DL channel. In our approach, the channel estimate for each narrowband resource block is readily available after identifying the DL channel parameters \cite{Han:2019, WMMSE-SAA, mao:twc:24}. Consequently, the precoder design algorithm is applied independently to each resource block, making it unnecessary to retain the sub-carrier index $m$ to define the DL channel.}
}}
Accordingly, assuming the difference between the UL carrier frequency and the target DL resource block is $f$, the DL channel as a function of $f$ is given by
\begin{align} \label{dlChannel}
    \boldsymbol{\mathbf{h}}_{k}(f) = \sum \limits^{L_k^{\mathrm{dl}}}_{\ell=1}\mathrm{\alpha}^{\mathrm{dl}}_{k, \ell} \boldsymbol{\mathrm{a}}\left(\theta^{\mathrm{dl}}_{k, \ell}; \lambda^{\mathrm{dl}} \right)e^{-j2\mathrm{\pi}f \tau^{\mathrm{dl}}_{k, \ell}} \in \mathbb{C}^{N \times 1},
\end{align}
where $\mathrm{\alpha}^{\mathrm{dl}}_{k, \ell}, \theta^{\mathrm{dl}}_{k, \ell}, \tau^{\mathrm{dl}}_{k, \ell}, L^{\mathrm{dl}}_k$ are defined in the same way as for the UL channel, and $\lambda^{\mathrm{dl}}$ denotes the wavelength of the corresponding resource block.

As presented in \cite{Deokhwan, Dina} and also supported by an actual measurement campaign \cite{T.Choi:2021}, 
the UL and DL channels share several frequency-invariant parameters. 
We summarize the partial reciprocity relationship for all $(k,\ell)$ as follows: 
\begin{itemize}
    \item Number of channel paths: $L_k^{\mathrm{ul}} = L_k^\mathrm{dl} \triangleq L_k $,
    \item AoA (or AoD): $\theta^{\mathrm{ul}}_{k, \ell} = \theta^{\mathrm{dl}}_{k, \ell}$,
    \item Propagation delay: $\tau^{\mathrm{ul}}_{k, \ell} = \tau^{\mathrm{dl}}_{k, \ell}$,
    \item Complex path gain: $\mathrm{\alpha}^{\mathrm{ul}}_{k, \ell} \approx \mathrm{\alpha}^{\mathrm{dl}}_{k, \ell}$.
\end{itemize}

We emphasize the critical role of frequency-invariant parameters in reconstructing the DL channel from UL training. However, the last assumption of frequency-invariant complex path gain remains controversial. 
Although \cite{Deokhwan, Dina} claimed that each path gain is also frequency-invariant, i.e. $\mathrm{\alpha}^{\mathrm{ul}}_{k, \ell} = \mathrm{\alpha}^{\mathrm{dl}}_{k, \ell}, \forall (k,\ell),$ the real-world measurement campaign \cite{Han:2019, Zhong:2020} reported that the instantaneous channel gains may differ. 
Instead, they tend to have a highly correlated second moment \cite{Zhong:2020}.
To account for this, we adopt the DL channel gain model using the first-order Gauss-Markov model:
\textcolor{black}{
\begin{align}\label{real_channel}
    {\mathrm{\alpha}^{\mathrm{dl}}_{k, \ell} = \eta_{k, \ell} \mathrm{\alpha}^{\mathrm{ul}}_{k, \ell} + \sqrt{1-\eta_{k, \ell}^2} \beta_{k, \ell}, \beta_{k, \ell} \sim \mathcal{CN}(0,\sigma^2_{\mathrm{path},k}), \forall \ell ,}
\end{align}
where $\eta_{k,\ell}$ is the correlation between $\ell$-th path of the DL and UL channels for user $k$.}
For example, if ${\eta_{k,\ell} = 1, \forall{(k, \ell)}}$, our model \eqref{real_channel} reduces to the case of perfect reciprocity, while ${\eta_{k,\ell}} = 0, \forall{(k, \ell)}$ indicates that UL/DL gains are independent, where our model includes \cite{Deokhwan, Dina} as an edge case and will allow us to investigate performance as a function of path correlation. 

\subsection{Signal Model}
We consider RSMA, which is a robust multiple access technique against imperfect CSIT \cite{Park:2023, Amor:2023}. 
In this paper, we employ $1$-layer RSMA so that only one layer of common messages for the all users is used \cite{Primer-RSMA}. \textcolor{black}{Specifically, the intended message $W_k$ for user $k$ is split into a common part $W_{\mathrm{c}, k}$ and a private part $W_{\mathrm{p}, k}$. Then, the common parts of entire users' messages are combined to yield a common message $W_{\mathrm{c}}$. Subsequently, the $K+1$ messages (the common message $W_{\mathrm{c}}$ and the $K$ private messages, $W_{\mathrm{p}, 1}, \ldots, W_{\mathrm{p}, K}$) are encoded into the symbols $s_{\mathrm{c}}, s_{\mathrm{p}, 1}, \ldots, s_{\mathrm{p}, K}$ respectively.
Note that the common message $s_{{\mathrm{c}}}$ is designed to be decodable to the all users, which implies that the common rate is not larger than the minimum common rate of entire users.}

\textcolor{black}{The transmit signal at the BS is then formed by combining the messages and linear precoding vectors as
\begin{align} 
{\mathbf{x}} ={\mathbf{f}}_{\mathrm{c}} s_{\mathrm{c}} + \sum _{k = 1}^{K}{\mathbf{f}}_{k} s_{k}, 
\end{align}
where ${\mathbf{f}}_{\mathrm{c}}, {\mathbf{f}}_{k} \in \mathbb{C}^{N\times 1}$ are precoding vectors for the common and private stream for user $k$, respectively. The transmit power constraint is $\left \|{{\mathbf{f}}_{\mathrm{c}} }\right \|^{2} + \sum _{k=1}^{K} \left \|{ {\mathbf{f}}_{k} }\right \|^{2} \le 1$. {In addition, we assume Gaussian signaling with power constraint $P$, i.e., $s_{\mathrm{c}},s_k \sim \mathcal{CN}(0,P)$.}}
At user $k \in [K]$, the received signal is 
\begin{align} 
\label{eq:r_k}
r_{k}(f) = &\ {\mathbf{h}}_{k}^{\sf H}(f) {\mathbf{f}}_{\mathrm{c}} s_{\mathrm{c}}+ {\mathbf{h}}_{k}^{\sf H}(f) {\mathbf{f}}_{k} s_{k} + \sum _{p = 1, p \neq k}^{K} {\mathbf{h}}_{k}^{\sf H}(f) {\mathbf{f}}_{p } s_{p } + z_{k},
\end{align}
where $z_{k} \sim \mathcal {CN}(0,\sigma ^{2})$ is additive white Gaussian noise (AWGN).

\section{DL Channel Reconstruction and Performance Characterization} \label{sec:recon}
\textcolor{black}{In this section, we explain a DL channel reconstruction method using the UL reference signal without CSI feedback. We then derive the achievable SE with the used DL channel reconstruction method.}



\subsection{DL Channel Reconstruction}
To begin, we define a vector $\mathbf{u}(\tau,\theta) \in \mathbb{C}^{MN\times1}$ as the received UL training signal across $M$ sub-carriers and $N$ antennas as follows:\textcolor{black}{
\begin{align}
    [{\mathbf{u}}(\tau,\theta)]_{m,n} = e^{-j2\pi\frac{nd}{\lambda^{\mathrm{ul}}_m}\sin\theta_{k,\ell}^{\mathrm{ul}}}e^{-j2\pi (\lfloor\frac{-M}{2}\rfloor+m-1) \triangle f \tau_{k, \ell} }\label{u_def},
\end{align}
where $[\cdot]_{m,n}$ denotes the element of the vector associated with the $m$-th sub-carrier and $n$-th antenna.}
Assuming an all-ones UL sounding reference signal is used, the received signal is given by 
\begin{align} 
{\mathbf{y}}_{k} = \sum \limits _{\ell = 1}^{L_k}{\alpha^{\mathrm{ ul}}_{k, \ell} {\mathbf{u}}(\tau^{\mathrm{ ul}} _{k, \ell},\theta^{\mathrm{ ul}} _{k, \ell})}+{\mathbf{w}}_k \in\mathbb{C}^{MN\times 1},
\label{eq:ul_sounding_rx}
\end{align} 
where ${\mathbf{w}}_k$ is AWGN. 
\textcolor{black}{
Our objective is to predict the DL channel $\mathbf{h}_k(f)$ using $\mathbf{y}_k$. 
To this end, we first obtain the UL triplet path parameters, i.e., $\left\{\left({\hat{\mathrm{\alpha}}}^{\mathrm{ul}}_{k, \ell}, {\hat{\tau}}^{\mathrm{ul}}_{k, \ell}, \hat{\theta}^{\mathrm{ul}}_{k, \ell}\right)\right\}_{\ell=1,...,{L}_k}$ from ${\bf{y}}_k$ and rebuild the DL channel $\hat{\mathbf{h}}_k(f)$ by incorporating the frequency invariance.
As demonstrated in \cite{drctnl:twc:18}, the number of channel paths in FDD massive MIMO is significantly smaller than the number of BS antennas, allowing us to formulate the UL channel parameter extraction problem as a CS problem. 
For solving this, we employ a variant of orthogonal matching pursuit, called 2D-NOMP. As demonstrated in \cite{Han:2019}, the 2D-NOMP algorithm has shown to be suitable for real-time hardware implementations.  
We explain the basic principle and detailed process of the 2D-NOMP as follows.}



\textcolor{black}{For every iteration, the 2D-NOMP algorithm finds the maximum likelihood (ML) estimate of the angle, delay, and channel gain of each path in UL, say ($\tilde{\alpha}_{k, \ell}, \tilde{\tau}_{k, \ell}, \tilde{\theta}_{k, \ell}$). 
With the estimate, the residual signal at the end of the $i$-th iteration is computed by
\begin{align} \label{eq:residual}
    \mathbf{y}_{\mathrm{r}} = \mathbf{y}_k - \sum^{i}_{\ell=1}\tilde{\alpha}_{k, \ell}\mathbf{u}(\tilde{\tau}_{k, \ell}, \tilde{\theta}_{k, \ell}),
\end{align}
and $\mathbf{y}_{\mathrm{r}}$ is initialized to $\mathbf{y}_k$.}
\textcolor{black}{The ML estimate of the parameters is attained by minimizing the residual power 
$\|\mathbf{y}_{\mathrm{r}} - \tilde{\alpha}_{k, i}\mathbf{u}(\tilde{\tau}_{k, i}, \tilde{\theta}_{k, i})\|^2$, 
which is equivalent to maximizing
\begin{align}\label{MLE}
        &f_{{\mathbf{y}}_\mathrm{r}}(\alpha_{k, \ell}, \tau_{k, \ell}, \theta_{k, \ell}) \nonumber \\ \triangleq& \ 2\mathrm{Re}\{\mathbf{y}_{\mathrm{r}}^{\sf H}\alpha_{k, \ell}\mathbf{u}(\tau_{k, \ell},\theta_{k, \ell})\}-|\alpha_{k, \ell}|^2\|\mathbf{u}(\tau_{k, \ell},\theta_{k, \ell})\|^2.
\end{align}}\noindent
Based on this, the three following steps are performed during the $i$-th iteration until a stopping criterion is met.

\textcolor{black}{
\textit{1. New detection}:
    To maximize \eqref{MLE}, the estimate for $i$-th path for user $k$, i.e., $(\tilde \tau_{k,i}, \tilde \theta_{k,i})$ is chosen from a predefined parameter grid, $\Omega$ as
    \begin{align}\label{basis-update}
        (\tilde{\tau}_{k,i}, \tilde{\theta}_{k,i}) = \argmax_{(\tau, \theta) \in \Omega}\frac{|\mathbf{u}^{\sf H}(\tau, \theta)\mathbf{y}_{\mathrm{r}}|^2}{\|\mathbf{u}(\tau, \theta)\|^2}.
    \end{align}
    With $(\tilde \tau_{k,i}, \tilde \theta_{k,i})$, we obtain an estimate of the channel gain as
    \begin{align}\label{gain-update}
        \tilde{\alpha}_{k,i} = \frac{\mathbf{u}^{\sf H}(\tilde \tau_{k,i}, \tilde \theta_{k,i})\mathbf{y}_{\mathrm{r}}}{\|\mathbf{u}(\tilde \tau_{k,i}, \tilde \theta_{k,i})\|^2}.
    \end{align}}
    
\textit{2. Newton refinement}: Since the parameter is chosen within a finite grid, $\Omega$, an off-grid error may occur as a result. Then, in pursuit of reducing the off-grid error, we define the Newton step and gradient step as 
\begin{align}
        &\mathbf{n}_{{\mathbf{y}}_{\mathrm{r}}}(\tilde{\alpha}, \tilde{\tau}, \tilde{\theta}) = -\nabla^2_{(\tau, \theta)}f^{-1}_{{\mathbf{y}}_{\mathrm{r}}}(\tilde{\alpha}, \tilde{\tau}, \tilde{\theta}) \nabla_{(\tau, \theta)}f_{{\mathbf{y}}_{\mathrm{r}}}(\tilde{\alpha}, \tilde{\tau}, \tilde{\theta}), \\
        &\mathbf{g}_{{\mathbf{y}}_{\mathrm{r}}}(\tilde{\alpha}, \tilde{\tau}, \tilde{\theta}) = \nabla_{({\tau}, {\theta})}f_{{\mathbf{y}}_{\mathrm{r}}}(\tilde{\alpha}, \tilde{\tau}, \tilde{\theta}).
\end{align}
We then refine $(\tilde{\tau}_{k,i}, \tilde{\theta}_{k,i})$ obtained in \eqref{basis-update} over the continumm via Newton's method \cite{NOMP}, i.e.,
\textcolor{black}{
\begin{align}\label{newton-step}
    &(\tilde{\tau}_{k,i}, \tilde{\theta}_{k,i}) =  (\tilde{\tau}_{k,i}, \tilde{\theta}_{k,i})\notag
    \\&+
    \begin{cases}
    \mathbf {n}_{ {\mathbf {y}}_\mathrm{r}}(\tilde{\alpha}_{k,i}, \tilde{\tau}_{k,i}, \tilde{\theta}_{k,i}), & \text{if } \nabla _{(\tau, \theta)}^{2}f_{{\mathbf {y}}_{\rm{r}}}(\tilde{\alpha}_{k,i}, \tilde{\tau}_{k,i}, \tilde{\theta}_{k,i})\prec \mathbf {0},\\
    \mu\mathbf {g}_{ {\mathbf {y}}_\mathrm{r}}(\tilde{\alpha}, \tilde{\tau}_{k,i}, \tilde{\theta}_{k,i}), & \text{else},
    \end{cases}
\end{align}}\noindent
where $\mu$ is the learning rate. After this refinement, the corresponding gain of the $i$-th path is recalculated as in \eqref{gain-update}.
We cyclically repeat this refinement process $R_c$ times for all the paths found up to this point. 

\textit{3. Update of gains}: In this step, by using the estimates $(\tilde{\tau}_{k, \ell},\tilde{\theta}_{k, \ell}), \forall \ell\in[i]$ obtained in the previous step, we lastly calibrate the gains of all the paths with least squares (LS): 
\begin{align}
    [\tilde{\alpha}_{k, 1}, \tilde{\alpha}_{k, 2}, \ldots,\tilde{\alpha}_{k, i}]=\tilde{\mathbf{U}}^{\dagger}\mathbf{y}_k,
\end{align}
where $\tilde{\mathbf{U}} = [\mathbf{u}(\tilde{\tau}_{k, 1}, \tilde{\theta}_{k, 1}), \mathbf{u}(\tilde{\tau}_{k, 2}, \tilde{\theta}_{k, 2}),\ldots,\mathbf{u}(\tilde{\tau}_{k, i}, \tilde{\theta}_{k, i})]$.

The algorithm terminates once a stopping criterion is met, often based on the false alarm rate. \textcolor{black}{We denote the final algorithm output as $\left\{\hat{\alpha}^{\mathrm{ul}}_{k, \ell}, \hat{\tau}^{\mathrm{ul}}_{k, \ell}, \hat{\theta}^{\mathrm{ul}}_{k, \ell}\right\}_{\ell=1,... ,{L}_k}$. 
Building on this, we let $\left\{\hat{\alpha}^{\mathrm{dl}}_{k, \ell}, \hat{\tau}^{\mathrm{dl}}_{k, \ell}, \hat{\theta}^{\mathrm{dl}}_{k, \ell}\right\}_{\ell=1,. ..,{L}_k} = \left\{\eta_{k,\ell} \hat{\alpha}^{\mathrm{ul}}_{k, \ell}, \hat{\tau}^{\mathrm{ul}}_{k, \ell}, \hat{\theta}^{\mathrm{ul}}_{k, \ell}\right\}_{\ell=1,... ,{L}_k}$ by incorporating the UL/DL frequency invariance property and leveraging the channel gain model in \eqref{real_channel}.
As shown \cite{Han:2019}, the 2D-NOMP algorithm achieves near-optimal mean squared error (MSE) performance, approaching the CRLB level.}




Finally, we reconstruct the DL channel as 
\begin{align} \label{estDlChannel}
    \boldsymbol{\hat{\mathbf{h}}}_{k}(f) = \sum \limits^{{L}_k}_{\ell=1}\hat{\mathrm{\alpha}}^{\mathrm{dl}}_{k, \ell}\boldsymbol{\mathrm{a}}\left(\hat{\theta}^{\mathrm{dl}}_{k, \ell}; \lambda^{\mathrm{dl}}\right)e^{-j2\mathrm{\pi}f \hat{\tau}^{\mathrm{dl}}_{k, \ell}}.
\end{align}
For conciseness, $\boldsymbol{{\mathbf{h}}}_{k}$ and $\hat{\boldsymbol{{\mathbf{h}}}}_{k}$ will represent the $\boldsymbol{{\mathbf{h}}}_{k}(f)$ and $\hat{\boldsymbol{{\mathbf{h}}}}_{k}(f)$, respectively, and $r_k$ will represent the $r_k(f)$ hereafter.
Before formulating the problem with the estimated channel, we briefly explain user grouping in the following remark.
\subsection{SE Characterization and Problem Formulation}
In this subsection, we characterize the SE 
under the assumption that the frequency difference between UL and DL is $f$. Then we formulate the main problem of this paper. \textcolor{black}{As shown in \cite{Park:2023, WMMSE-SAA}, the exact form of the instantaneous SE for both the common and private messages in each fading block, given imperfect CSIT, cannot be expressed in closed form. To resolve this, we derive a useful lower bound on the SE.}
Let us model the relationship between the true and reconstructed channel as ${\mathbf{h}}_{k} = \hat{\mathbf{h}}_{k} + {\mathbf{e}}_{k}$, where ${\mathbf{e}}_{k}$ represents the DL channel reconstruction error.
Then, we can express the average power of the received signal at user \( k \) in \eqref{eq:r_k}, normalized by \( P \), (i.e., \( \frac{\mathbb{E}\big\{|r_{k}|^{2}\big\}}{P}) \) as
\begin{align}
\overbrace{|\hat{\mathbf{h}}_{k}^{\sf{H}}\mathbf{f}_{\mathrm{c}}|^{2}}^{S_{\mathrm{c}}} + \underbrace{\overbrace{|\hat{\mathbf{h}}_{k}^{\sf{H}}\mathbf{f}_{k}|^{2}}^{S_{k}}+
\overbrace{\sum_{p\neq k}^K |\hat{\mathbf{h}}_{k}^{\sf{H}}\mathbf{f}_{p}|^{2}+|\mathbf{e}_{k}^{\sf{H}}\mathbf{f}_{\mathrm{c}}|^{2} + \sum_{p=1}^K |\mathbf{e}_{k}^{\sf{H}}\mathbf{f}_{p}|^{2} 
+ \frac{\sigma^{2}}{P}.}^{I_k}}_{I_{\mathrm{c}}}
\end{align}
Subsequently, by treating ${\mathbf{e}}_{k}$ as independent Gaussian noise, we characterize a lower bound on the instantaneous SE of $s_{\mathrm{c}}$ as 
\begin{align} 
&R_{\mathrm{c}}^{\sf ins.}(k) \mathop {\ge}^{(a)}\mathbb {E}_{\{{\mathbf{e}_k|\hat{\boldsymbol{\psi}}_k}\}} \left [{ \log _{2} \left (1+S_{\mathrm{c}}I_{\mathrm{c}}^{-1}\right) }\right] \\
&\mathop {\ge}^{(b)}\log _{2} \left ({1 + \frac {|\hat {\mathbf{h}}_{k}^{\sf H} {\mathbf{f}}_{\mathrm{c}}|^{2}}{\begin{Bmatrix} {\sum _{p = 1}^{K} |\hat {\mathbf{h}}_{k}^{\sf H} {\mathbf{f}}_{p} |^{2} + {\mathbf{f}}_{\mathrm{c}}^{\sf H } \mathbb {E} \left [{ {\mathbf{e}}_{k} {\mathbf{e}}_{k}^{\sf H} }\right] {\mathbf{f}}_{\mathrm{c}}} \\  + {\sum _{p = 1}^{K} {\mathbf{f}}_{p }^{\sf H} \mathbb {E} \left [{{\mathbf{e}}_{k} {\mathbf{e}}_{k}^{\sf H} }\right] {\mathbf{f}}_{p}} {+ \frac {\sigma ^{2}}{P}} \end{Bmatrix} } }\right) \label{R_c_up} \\
&\mathop {=}^{(c)}\log _{2} \left ({1 + \frac {|\hat {\mathbf{h}}_{k}^{\sf H} {\mathbf{f}}_{\mathrm{c}}|^{2}}{\begin{Bmatrix} {\sum _{p = 1}^{K} |\hat {\mathbf{h}}_{k}^{\sf H} {\mathbf{f}}_{p} |^{2} + {\mathbf{f}}_{\mathrm{c}}^{\sf H} {\boldsymbol{\Phi }}_{k} {\mathbf{f}}_{\mathrm{c}}} \\ { + \sum _{p = 1}^{K} {\mathbf{f}}_{p }^{\sf H} {\boldsymbol{\Phi }}_{k} {\mathbf{f}}_{p } + \frac {\sigma ^{2}}{P}} \end{Bmatrix} } }\right)  \label{R_c} \\
&\triangleq  R_{\mathrm{c}}^{\sf lb}(k). \nonumber 
\end{align}
where $(a)$ follows from treating the DL channel reconstruction error as independent Gaussian noise \textcolor{black}{which gives the worst case mutual information \cite{Park:2023}, 
$(b)$ follows Jensen's inequality, and $(c)$ comes from $\mathbb {E}_{\{{\mathbf{e}_k|\hat{\boldsymbol{\psi}}_k}\}}[{\bf{e}}_k {\bf{e}}_k^{\sf H}] = {\boldsymbol{\Phi }}_{k} $, where we omit the notation of ${\{{\mathbf{e}_k|\hat{\boldsymbol{\psi}}_k}\}} $ due to the space limitation.}
To guarantee the decodability of $s_{{\mathrm{c}}}$ for the users in $[K]$, the code rate of $s_{{\mathrm{c}}}$ is determined as
$\min _{k \in [K]} \{R_{\mathrm{c}}^{\sf lb}(k)\} $. 
Provided that the proper code rate is used, it is guaranteed that $s_{{\mathrm{c}}}$ is successfully decoded and eliminated with SIC.
After SIC, we derive a lower bound on the instantaneous SE of $s_k$ using the similar process as in \eqref{R_c}:
\begin{align} 
&R_{k}^{\sf ins.} \mathop \geq \mathbb {E}_{\{{\mathbf{e}_k|\hat{\boldsymbol{\psi}}_k}\}} \left [{ \log _{2} \left (1+S_{k}I_{k}^{-1}\right) }\right] \\&\mathop {\ge }^{}\log _{2} \!\left ({\!1 \!+ \!\frac {|\hat {\mathbf{h}}_{k}^{\sf H} {\mathbf{f}}_{k}|^{2}} {\begin{Bmatrix} {\sum _{p = 1, p \neq k}^{K} |\hat {\mathbf{h}}_{k}^{\sf H} {\mathbf{f}}_{p} |^{2}}  { + \sum _{p = 1}^{K} {\mathbf{f}}_{p }^{\sf H} {\boldsymbol{\Phi }}_{k} {\mathbf{f}}_{p } + \frac {\sigma ^{2}}{P}} \end{Bmatrix} }\!}\right) \label{R_k} \\
&\triangleq R_{k}^{\sf lb}. \nonumber 
\end{align}
Leveraging the derived SE expressions, we formulate the sum SE maximization problem. 
As explained above, to ensure the decodability of $s_{{\mathrm{c}}}$, the code rate of $s_{{\mathrm{c}}}$ is set as $\min _{k \in [K]} \{ R_{\mathrm{c}}^{\sf lb}(k)\} $. With this, the sum SE maximization problem is given by 
\begin{align}
\mathscr{P}_1: &\mathop{{\text {maximize}}}_{{\mathbf{f}}_{\mathrm{c}},{\mathbf{f}}_{1}, \cdots, {\mathbf{f}}_{K}} \;\; \min _{k \in [K]} \{ R_{\mathrm{c}}^{\sf lb}(k)\} + \sum _{k = 1}^{K}  R_{k}^{\sf lb} \label{original} 
\\&{\text {subject to}} \;\; \left \|{ {\mathbf{f}}_{\mathrm{c}} }\right \|^{2} + \sum _{k = 1}^{K} \left \|{ {\mathbf{f}}_{k} }\right \|^{2} \le 1. \label{constraint}
\end{align}
Unfortunately, it is infeasible to directly solve \eqref{original} due non-smoothness and non-convexity entailed in the objective function.  
In the next subsection, we present our approach for reformulating \eqref{original} into a tractable form. 


\subsection{Problem Reformulation}
To tackle our main problem of $\mathscr{P}_1$, we first transform the objective function in \eqref{original} into a matrix form. To do this, we first define an unified precoding vector by stacking all precoders, ${\mathbf{f}}_{\mathrm{c}}, {\mathbf{f}}_{1}, \cdots, {\mathbf{f}}_{K}$ as
\begin{align} 
\bar {\mathbf{f}} = [{\mathbf{f}}_{\mathrm{c}}^{\sf T}, {\mathbf{f}}_{1}^{\sf T}, \cdots, {\mathbf{f}}_{K}^{\sf T}]^{\sf T} \in \mathbb {C}^{N(K+1) \times 1}. 
\end{align}
With $\bar {\bf{f}}$, we express the SINR of $s_{\mathrm{c}}$ as a form of Rayleigh quotient, i.e. ${\bar{\mathbf{f}}^{\sf H} \mathbf{A}_{\mathrm{c}}(k) \bar{\mathbf{f}}}/{\bar{\mathbf{f}}^{\sf H} \mathbf{B}_{\mathrm{c}}(k) \bar{\mathbf{f}}}$, where 
\begin{align}\label{A_B_common}
{\mathbf{A}}_{\mathrm{c}}(k) = & \, {\mathrm{diag}}\big(\overbrace{(\hat{\mathbf{h}}_{k} \hat{\mathbf{h}}_{k}^{\sf H} + {\boldsymbol{\Phi}}_{k}), \ldots, (\hat{\mathbf{h}}_{k} \hat{\mathbf{h}}_{k}^{\sf H} + {\boldsymbol{\Phi}}_{k})}^{(K+1)\ \mathrm{blocks}}\big) \nonumber  \\
& + {\mathbf{I}}_{N(K+1)}\frac{\sigma^2}{P}, \\
{\mathbf{B}}_{\mathrm{c}}(k) = & \, {\mathrm{diag}}\big({\boldsymbol{\Phi}}_{k},\overbrace{(\hat{\mathbf{h}}_{k} \hat{\mathbf{h}}_{k}^{\sf H} + {\boldsymbol{\Phi}}_{k}), \ldots, (\hat{\mathbf{h}}_{k} \hat{\mathbf{h}}_{k}^{\sf H} + {\boldsymbol{\Phi}}_{k})}^{K\ \mathrm{blocks}}\big)\nonumber  \\
& + {\mathbf{I}}_{N(K+1)}\frac{\sigma^2}{P}.
\end{align}
Similarly, the SINR of $s_k$ is written as ${\bar{\mathbf{f}}^{\sf H} \mathbf{A}_k \bar{\mathbf{f}}}/{\bar{\mathbf{f}}^{\sf H} \mathbf{B}_k \bar{\mathbf{f}}}$, where 
\begin{align}
{\mathbf{A}}_{k} = & \,  {\mathrm{diag}}  \big(\mathbf{0},  \overbrace{(\hat{\mathbf{h}}_{k} \hat{\mathbf{h}}_{k}^{\sf H} + {\boldsymbol{\Phi}}_{k}),\ldots, (\hat{\mathbf{h}}_{k} \hat{\mathbf{h}}_{k}^{\sf H} + {\boldsymbol{\Phi}}_{k})}^{K\ \mathrm{blocks}}\big) \nonumber \\
& + {\mathbf{I}}_{N(K+1)}\frac{\sigma^2}{P}, \\ \label{A_B_private}
{\mathbf{B}}_{k} = & \, {\mathrm{diag}}  \big(\mathbf{0}, \overbrace{(\hat{\mathbf{h}}_{k} \hat{\mathbf{h}}_{k}^{\sf H} + {\boldsymbol{\Phi}}_{k}),  \ldots,(\hat{\mathbf{h}}_{k} \hat{\mathbf{h}}_{k}^{\sf H} + {\boldsymbol{\Phi}}_{k})}^{(k-1)\ \mathrm{blocks}}, \nonumber \\ &\overbrace{\!\boldsymbol{\Phi}_k\!}^{(k+1)\text{-th block}},\overbrace{(\hat{\mathbf{h}}_{k} \hat{\mathbf{h}}_{k}^{\sf H} + {\boldsymbol{\Phi}}_{k}), \ldots, (\hat{\mathbf{h}}_{k} \hat{\mathbf{h}}_{k}^{\sf H} + {\boldsymbol{\Phi}}_{k})}^{(K-k)\ \mathrm{blocks}}\big) \nonumber \\
& + {\mathbf{I}}_{N(K+1)}\frac{\sigma^2}{P}.
\end{align}
With this reformulation, we exploit the LogSumExp technique to resolve the non-smoothness involved in \eqref{original}.
We approximate the non-smooth minimum function as 
\begin{align}
    \label{eq:logsumexp}
    \min_{i = 1,...,N}\{x_i\}  &\approx -\alpha \log\left(\frac{1}{N}  \sum_{i = 1}^{N} \exp\left( \frac{x_i}{-\alpha}  \right)\right) \\
    & \triangleq g(\{x_i\}_{i \in [N]}),
\end{align}
where $\alpha$ determines the accuracy of the approximation. 
As $\alpha \rightarrow 0$, the approximation becomes tight. 
Leveraging \eqref{eq:logsumexp}, we approximate the SE of $s_{{\mathrm{c}}}$ as 
\begin{align} \label{eq:approximation_R}
\min_{k \in [K]} \{ R_{{\mathrm{c}}}^{\sf lb}(k)\} \approx  g\left( \left\{ \frac{\bar{\mathbf{f}}^{\sf H} \mathbf{A}_{\mathrm{c}}(k) \bar{\mathbf{f}}}{\bar{\mathbf{f}}^{\sf H} \mathbf{B}_{\mathrm{c}}(k) \bar{\mathbf{f}}} \right\}_{k \in [K]} \right).
\end{align}
Using this, we reformulate the original problem $\mathscr{P}_1$ defined in \eqref{original} as 
\begin{align}
\mathscr{P}_2:\mathop {\text{maximize}}_{\bar{\mathbf{f}}} \  g\left( \left\{ \frac{\bar{\mathbf{f}}^{\sf H} \mathbf{A}_{\mathrm{c}}(k) \bar{\mathbf{f}}}{\bar{\mathbf{f}}^{\sf H} \mathbf{B}_{\mathrm{c}}(k) \bar{\mathbf{f}}} \right\}_{k \in [K]} \right) + \sum_{k = 1}^{K} \log_2 \left( \frac{\bar{\mathbf{f}}^{\sf H} \mathbf{A}_k \bar{\mathbf{f}}}{\bar{\mathbf{f}}^{\sf H} \mathbf{B}_k \bar{\mathbf{f}}} \right) \label{subProblem}.
\end{align}
In $\mathscr{P}_2$, we drop the constraint \eqref{constraint} since the objective function of $\mathscr{P}_2$ is scale invariant, which does not affect optimality \cite{Park:2023}. Even with this reformulation, however, it is still infeasible to solve $\mathscr{P}_2$, since the ECM ${\bf{\Phi}}_k$, $k \in [K]$ has not yet been determined. To address this, we estimate the ECM ${\bf{\Phi}}_k$, $k \in [K]$ in the next section.

\section{Error Covariance Matrix Estimation}
To understand the difficulty of obtaining an ECM without CSI feedback, we first summarize the conventional approaches.
For simplicity, we let $L_k = 1, \kappa_k = 0$, and $\sigma^2_{\mathrm{path}, k} = 1$ in the following explanation.  



\begin{itemize}
    \item {\bf{Bayesian approach}}: In this approach, we calculate the ECM by using known channel statistics. Assuming that the statistics of AoD, denoted by $\theta_k^{\rm{dl}}$, is known at the BS, then we obtain the channel covariance $\mathbb{E}[{\bf{h}}_k {\bf{h}}_k^{\sf H}] = \boldsymbol{\mathrm{a}}\left(\theta^{\mathrm{dl}}_{k}; \lambda^{\mathrm{dl}}\right) \cdot \boldsymbol{\mathrm{a}}^{\sf H}\left(\theta^{\mathrm{dl}}_{k}; \lambda^{\mathrm{dl}}\right) =  {\bf{R}}_k$. Assuming that linear MMSE estimation \cite{mao:twc:24} is used, the ECM is given as 
    \begin{align}
        \hat{{\bf{\Phi}}}_k = {\bf{R}}_k - {\bf{R}}_k \left ({\bf{R}}_k + \rho_{\rm{pilot}}^2 {\bf{I}}_N \right)^{-1} {\bf{R}}_k, \label{eq:closedECM}
    \end{align}
    where $\rho_{\rm{pilot}}^2 $ represents the inverse of the effective SNR of the pilot signals. 
    \item {\bf{Frequentist approach}}: In this case, instead of using the channel statistics, we assume that the observed error samples are given. Denoting the $i$-th DL channel error sample as ${\bf{e}}_k(i)$ with $i \in [T]$, the sample ECM is obtained as 
    \begin{align}
        \hat {\bf{\Phi}}_k = \frac{1}{T} \sum_{i = 1}^{T} {\bf{e}}_k(i) {\bf{e}}_k^{\sf H}(i) 
 +\rho {\bf{I}}, 
    \end{align}
    where $\rho$ is a regularization parameter to well-condition $\hat {\bf{\Phi}}_k$ \cite{covApprox}.
\end{itemize}
Unfortunately, neither approach is suitable in our case.  
First, we assume that no channel statistics are known at the BS. Further, it is difficult to analyze the ECM by using 2D-NOMP due to its non-linearity. 
Additionally, the sample ECM cannot be obtained in our scenario \cite{Deokhwan}.


To address this challenge, 
we turn our attention to 2D-NOMP's near-CRLB MSE performance as shown in \cite{NOMP}. This result implies that the ECM can be tightly estimated by leveraging the CRLB \cite{Rottenberg:2020}, whose trace gives a lower bound on the MSE. Considering the relationship between the UL and DL channels and assuming reciprocal path gains in \eqref{real_channel}, we derive the CRLB \cite{kay1993statistical, Rottenberg:2020} as 
\begin{align} \label{CRLB}
\boldsymbol{\Phi }_{k} \succcurlyeq \boldsymbol{\mathbf{J}}_k^{\sf{H}}(f)  \boldsymbol{\mathbf{I}}^{-1}(\boldsymbol{\psi}_k)\boldsymbol{\mathbf{J}}_k(f) = {\mathbf{C}}(f),
\end{align}
where $\boldsymbol{\psi}_k$ is the UL channel parameter vector defined as 
\begin{align}
\boldsymbol{\psi}_k &= (\bar{\psi}_{k,1}^{\sf{T}}, \ldots, \bar{\psi}_{k, L_k}^{\sf T})^{\sf T} \in \mathbb{R}^{4L_k\times1}, \\
\bar{\psi}_{k, \ell} &= (\theta_{k, \ell}^{\rm ul}, \tau_{k, \ell}^{\mathrm{ul}}, \mathrm{Re}\{\alpha_{k, \ell}^{\mathrm{ul}}\}, \mathrm{Im}\{\alpha_{k, \ell}^{\mathrm{ul}}\})^{\sf{T}} \in \mathbb{R}^{4\times1}.
\end{align}
Here, $\boldsymbol{\mathbf{I}}(\boldsymbol{\psi}_k) \in \mathbb{C}^{4L_k\times 4L_k}$ indicates the FIM of user $k$ and $\boldsymbol{\mathbf{J}}_k(f) \in \mathbb{C}^{4L_k\times N}$ denotes the Jacobian matrix of user $k$, evaluated at the frequency difference $f$ defined as 
\begin{align} 
    \boldsymbol{\mathbf{J}}_k(f) \triangleq \frac{\partial\mathbf{h}_k^{\sf{T}}(f)}{\partial\boldsymbol{\psi}_k},
\end{align}
where $\mathbf{h}_k(f)$ is in \eqref{dlChannel}. 



An implication of \eqref{CRLB} is that each diagonal element follows $[\boldsymbol{\Phi}_k]_{n,n}\geq[\mathbf{C}(f)]_{n,n}, \forall n$. \textcolor{black}{Given that 2D-NOMP exhibits the near-optimal MSE performance \cite{NOMP}, the MSE of 2D-NOMP can be closely approximated by the trace of CRLB, i.e., $\mathbb{E}\left[\|\mathbf{h}_k(f)-\hat{\mathbf{h}}_k(f)\|^2\right] = \mathrm{tr}\{\boldsymbol{\Phi}_k\}\approx\mathrm{tr}\{\mathbf{C}(f)\}$.}
Building on this, we approximate the ECM as a diagonal matrix by using the CRLB, i.e.,
\begin{align}\label{ECM_firstStep}
    {\boldsymbol{\Phi}}_k\approx\mathbf{C}(f)\circ\mathbf{I}_N.
\end{align}
To use \eqref{ECM_firstStep} as the estimated ECM\footnote{We note that our ECM estimation depends only on the diagonal components since its purpose is based on the 2D NOMP algorithm's near-CRLB performance. However, in Section \ref{sec:sim}, we will show that this approach is still very useful, as the DL channel reconstruction errors are not negligible.}, however, we note that 
the Jacobian and FIM, i.e., $\mathbf{J}_k(f), {\bf{I}}(\boldsymbol{\psi}_k)$ should even be evaluated using the true UL channel parameter $\boldsymbol{\psi}_k$. This is typically not available in practice \cite{efron1978assessing}. 
One maybe tempted to use the estimated parameter $\hat{\boldsymbol{\psi}}_k$ to calculate \eqref{ECM_firstStep}, yet this compromises the interpretation of the FIM. 
To address this, we propose to use the O-FIM \cite{observedFisher}.
The relation of the O-FIM to the FIM and derivation of the O-FIM are provided in the following remark and lemma, respectively. 
\begin{remark} [O-FIM and FIM]
\normalfont 
By definition, the FIM is 
\begin{align} \label{FIM}
    \boldsymbol{\mathbf{I}}(\boldsymbol{\psi}_k) = \mathbb{E}\left[-\frac{\partial^2 \log f(\mathbf{y}|\boldsymbol{\psi}_k)}{\partial \boldsymbol{\psi}_k \partial \boldsymbol{\psi}_k^{\sf T}}
    \right],
\end{align}
where the expectation is taken with respect to the observation $\mathbf{y}$ conditioned on the true parameter $\boldsymbol{\psi}_k$. On the contrary, the O-FIM is defined as an instantaneous observation of the FIM. This observed information can be interpreted as a sampled version of \eqref{FIM}, given by 
\begin{align}\label{O-FIM}
    \tilde{\boldsymbol{\mathbf{I}}}(\hat{\boldsymbol{\psi}}_k) = -\frac{\partial^2\log f(\mathbf{y}|\boldsymbol{\psi}_k)}{\partial\boldsymbol{\psi}_k \partial\boldsymbol{\psi}_k^{\sf T}}\Bigg|_{\boldsymbol{\psi}_k=\hat{\boldsymbol{\psi}}_k},
\end{align}
which can be computed straightforwardly with the estimated parameters.
\end{remark}
\begin{lemma} \label{first lemma}
Assuming that the likelihood of the observed signal $\mathbf{y}$ follows Gaussian distribution, each element of the O-FIM 
in \eqref{O-FIM}, calculated only using the UL estimated parameter, $\hat{\boldsymbol{\psi}}_k$, is 
\begin{align}
\label{Observed Fisher}
[\tilde{\boldsymbol{\mathbf{I}}}(\hat{\boldsymbol{\psi}}_k)]_{u, v} = \frac{2}{\sigma^2} &\mathrm{Re} 
 \left\{\sum_{n=1}^N\sum_{m=1}^M\left(\frac{\partial\bar{\mathbf{y}}^{*}_{n,m}}{\partial\psi_u}\frac{\partial\bar{\mathbf{y}}_{n,m}}{\partial\psi_v}\right) \right. \nonumber \\
 \quad &-\left.\sum_{n=1}^N\sum_{m=1}^M (\mathbf{y}_{n,m}-\bar{\mathbf{y}}_{n,m})^{*}\frac{\partial^2\bar{\mathbf{y}}_{n,m}}{\partial\psi_u\partial\psi_v}
 \right\}\Bigg|_{\boldsymbol{\psi}_k = \hat{\boldsymbol{\psi}}_k},
\end{align}
where $\bar {\bf{y}}$ is the reconstructed UL channel with the parameter set $\boldsymbol{\psi}_k$ defined as 
\begin{align}
\bar{\mathbf{y}} &\triangleq \sum_{\ell=1}^{L_k}\alpha_{k, \ell}^{\mathrm{ul}}\mathbf{u}(\tau^{\mathrm{ul}}_{k, \ell}, \theta^{\mathrm{ul}}_{k, \ell}) \in \mathbb{C}^{MN\times1}.
\end{align}
\end{lemma}
\begin{proof}
    See Appendix A. 
\end{proof}
\textcolor{black}{
In \cite{efron1978assessing, observedFisher}, it was found that the inverse of the O-FIM serves as a sampled version of the CRLB and is optimal for estimating the realized squared error. That is to say, the inverse of the O-FIM provides the best prediction of the realized error based on the given received signal.} 
Namely, with the Jacobian transformation \cite{Rottenberg:2020}, we have 
\begin{align}\label{OFIM-property}
\tilde{\boldsymbol{\mathbf{J}}}_k^{\sf{H}}(f)  &\tilde{\boldsymbol{\mathbf{I}}}^{-1}(\hat{\boldsymbol{\psi}}_k)\tilde{\boldsymbol{\mathbf{J}}}_k(f) \notag \\
&= \argmin_{\hat{\boldsymbol{\Phi}}_k} \ \mathbb{E}_{\{\mathbf{e}_k|\hat{\boldsymbol{\psi}}_k\}}\left[\sum_{m=1}^{N}\left[\mathbf{e}_k\mathbf{e}_k^{\sf H}-\hat{\boldsymbol{\Phi}}_k\right]_{m,m}^2\right],
\end{align}
where $\tilde{\boldsymbol{\mathbf{J}}}(f)$ is the Jacobian matrix with ${\boldsymbol{\psi}}_k$ at $f$ and O-FIM 
$\tilde{\boldsymbol{\mathbf{I}}}(\hat{\boldsymbol{\psi}}_k)$ is obtained from Lemma \ref{first lemma}.
Applying Jensen's inequality on the right-hand side of \eqref{OFIM-property} and omitting the notation ${\{{\mathbf{e}_k|\hat{\boldsymbol{\psi}}_k}\}}$, we obtain
\begin{align}\label{OFIM-ECM}
\sum_{m=1}^{N}\left[\mathbb{E}\left[\mathbf{e}_k\mathbf{e}_k^{\sf H}\right] - \hat{\boldsymbol{\Phi}}_k \right]_{m,m}^2\leq\mathbb{E}\left[\sum_{m=1}^{N}\left[\mathbf{e}_k\mathbf{e}_k^{\sf H}-\hat{\boldsymbol{\Phi}}_k\right]_{m,m}^2\right].
\end{align}
\textcolor{black}{Note that our aim is to estimate the ECM, i.e., $\mathbb{E}[\mathbf{e}_k\mathbf{e}^{\sf H}_k]$, using the matrix $\hat{\mathbf{\Phi}}_k$. The sum of its diagonal squared errors is represented on the left-hand side of \eqref{OFIM-ECM}. 
As shown in \eqref{OFIM-ECM}, our solution derived from \eqref{OFIM-property} in fact minimizes an upper bound on the sum of the squared errors in the diagonal elements for the ECM estimation.}
This supports the use of the O-FIM to estimate diagonal elements of the ECM. 
We summarize this in the following proposition. 
\begin{proposition} \label{prop ofim}
The estimated ECM represented in a diagonal form, which minimizes the upper bound of the MSE, can be derived using the O-FIM as follows:
\begin{align}\label{approx-w/o-disc}
\hat{\boldsymbol{\Phi}}_k
=
     \left(\tilde{\boldsymbol{\mathbf{J}}}_k^{\sf{H}}(f) \tilde{\boldsymbol{\mathbf{I}}}^{-1}(\hat{\boldsymbol{\psi}}_k)\tilde{\boldsymbol{\mathbf{J}}}_k(f)\right)\circ\mathbf{I}_N \triangleq \tilde{\mathbf{C}}(f).
\end{align}
\end{proposition}
We note that $\hat{\boldsymbol{\Phi}}_k$ in Proposition \ref{prop ofim} can be calculated solely by the estimated UL channel parameter $\hat{\boldsymbol{\psi}}_k$, which is valuable in practical scenarios.
However, it is worth mentioning that $\hat{\boldsymbol{\Phi}}_k$ in Proposition \ref{prop ofim} assumes a reciprocal path gain, i.e., $\textcolor{black}{\eta_{k,\ell} = 1, \forall{k, \ell}}$ \cite{Rottenberg:2020}. It is necessary to further calibrate the estimated ECM for a case where \textcolor{black}{$0 \leq\eta_{k,\ell} \leq 1, \forall{k, \ell}$}. 
We obtain this in the following corollary.
\begin{corollary}\label{corollary}
\textcolor{black}{For general $\eta_{k, \ell}$, the estimated ECM can be generalized to 
\begin{align}\label{eq:phi_final}
\hat{\boldsymbol{\Phi}}_k = \left(\sum_{\ell=1}^{L_k}\frac{\eta^2_{k,\ell}}{L_k}\right)\tilde{\mathbf{C}}(f)+\left(\sum_{\ell=1}^{L_k}\frac{1-\eta_{k,\ell}^2}{L_k}\right)\mathbf{I}_{n},
\end{align}
where $\tilde{\mathbf{C}}(f)$ is given by \eqref{approx-w/o-disc}.}
\end{corollary}
\begin{proof}
    See Appendix B.
\end{proof}
Using $\hat{\boldsymbol{\Phi}}_k$ in \eqref{eq:phi_final}, 
we eventually reach $\mathscr{P}_3$ as follows: 
\begin{align}
\mathscr{P}_3:\mathop {\text{maximize}}_{\bar{\mathbf{f}}} \ g\left( \left\{ \frac{\bar{\mathbf{f}}^{\sf H} \hat{\mathbf{A}}_{\mathrm{c}}(k) \bar{\mathbf{f}}}{\bar{\mathbf{f}}^{\sf H} \hat{\mathbf{B}}_{\mathrm{c}}(k) \bar{\mathbf{f}}} \right\}_{k \in [K]} \right) + \sum_{k = 1}^{K} \log_2 \left( \frac{\bar{\mathbf{f}}^{\sf H} \hat{\mathbf{A}}_k \bar{\mathbf{f}}}{\bar{\mathbf{f}}^{\sf H} \mathbf{B}_k \bar{\mathbf{f}}} \right) \label{mainProblem},
\end{align}
where we define $\hat{\mathbf{A}}_{\mathrm{c}}(k), \hat{\mathbf{B}}_{\mathrm{c}}(k), \hat{\mathbf{A}}_{k}$, and $\hat{\mathbf{B}}_{k}$ by replacing every occurrence of $\boldsymbol{\Phi}_k$ in \eqref{A_B_common} through \eqref{A_B_private} with $\hat{\boldsymbol{\Phi}}_k$.
Note that every element in \eqref{mainProblem} can be configured during UL training. Now, we are ready to develop an algorithm to solve \eqref{mainProblem}. 

\section{Precoder Optimization}

In this section, we propose a precoding optimization method to solve problem \eqref{mainProblem}. 
Taking inspiration from \cite{Park:2023, Jiwook}, the proposed method is centered on the following first-order optimality condition. 
\begin{theorem} \label{theo:optimal}
For problem \eqref{mainProblem}, the first-order optimality condition is satisfied when the following holds:
\begin{align} {\mathbf{B}}_{\sf KKT}^{-1} (\bar {\mathbf{f}}){\mathbf{A}}_{\sf KKT} (\bar {\mathbf{f}}) \bar {\mathbf{f}} = \lambda (\bar {\mathbf{f}}) \bar {\mathbf{f}},\label{KKT condition}\end{align}
where each matrix is shown below. 
\begin{align} &{\mathbf{A}}_{\sf KKT}(\bar {\mathbf{f}})=\lambda _{\sf num} (\bar {\mathbf{f}}) \times \Bigg [{ \sum _{k = 1}^{K} \frac {\hat{\mathbf{A}}_{k}}{\bar {\mathbf{f}}^{\sf H} \hat{\mathbf{A}}_{k} \bar {\mathbf{f}}} } \notag \\
&\qquad{  {  + \sum _{k \in [K]}\left ({\frac {\exp \left ({\frac {1}{-\alpha } \frac {\bar {\mathbf{f}}^{\sf H} \hat{\mathbf{A}}_{\mathrm{c}}(k) \bar {\mathbf{f}}}{\bar {\mathbf{f}}^{\sf H} \hat{\mathbf{B}}_{\mathrm{c}}(k) \bar {\mathbf{f}}} }\right)}{\sum _{j \in [K]} \exp \left ({\frac {1}{-\alpha } \log _{2} \left ({\frac {\bar {\mathbf{f}}^{\sf H} \hat{\mathbf{A}}_{\mathrm{c}}(j) \bar {\mathbf{f}}}{\bar {\mathbf{f}}^{\sf H} \hat{\mathbf{B}}_{\mathrm{c}}(j) \bar {\mathbf{f}}} }\right) }\right) } \frac {\hat{\mathbf{A}}_{\mathrm{c}} (k)}{\bar {\mathbf{f}}^{\sf H} \hat{\mathbf{A}}_{\mathrm{c}} (k)~\bar {\mathbf{f}} } }\right) }}\Bigg], \label{A_KKT}
\end{align}
\begin{align}
&{\mathbf{B}}_{\sf KKT}(\bar {\mathbf{f}})=\lambda _{\sf den} (\bar {\mathbf{f}}) \times \Bigg [{ \sum _{k = 1}^{K} \frac {\hat{\mathbf{B}}_{k}}{\bar {\mathbf{f}}^{\sf H} \hat{\mathbf{B}}_{k} \bar {\mathbf{f}}} } \notag \\
&\qquad{  {  + \sum _{k \in [K]}\left ({\frac {\exp \left ({\frac {1}{-\alpha } \frac {\bar {\mathbf{f}}^{\sf H} \hat{\mathbf{A}}_{\mathrm{c}}(k) \bar {\mathbf{f}}}{\bar {\mathbf{f}}^{\sf H} \hat{\mathbf{B}}_{\mathrm{c},}(k) \bar {\mathbf{f}}} }\right)}{\sum _{j \in [K]} \exp \left ({\frac {1}{-\alpha } \log _{2} \left ({\frac {\bar {\mathbf{f}}^{\sf H} \hat{\mathbf{A}}_{\mathrm{c}}(j) \bar {\mathbf{f}}}{\bar {\mathbf{f}}^{\sf H} \hat{\mathbf{B}}_{\mathrm{c}}(j) \bar {\mathbf{f}}} }\right) }\right) } \frac {\hat{\mathbf{B}}_{\mathrm{c}} (k)~}{\bar {\mathbf{f}}^{\sf H} \hat{\mathbf{B}}_{\mathrm{c}} (k)~\bar {\mathbf{f}} } }\right) }}\Bigg], \label{B_KKT} \end{align}
\begin{align}
&\lambda (\bar {\mathbf{f}}) =\prod _{k = 1}^{K} \left ({\frac {\bar {\mathbf{f}}^{\sf H} \hat{\mathbf{A}}_{k} \bar {\mathbf{f}}}{\bar {\mathbf{f}}^{\sf H} \hat{\mathbf{B}}_{k} \bar {\mathbf{f}}} }\right) \notag \\
&\qquad \times \, \left \{{\frac {1}{K} \sum _{k \in [K]} \exp \!\left ({\!\log _{2} \left ({\frac {\bar {\mathbf{f}}^{\sf H} \hat{\mathbf{A}}_{\mathrm{c}}(k) \bar {\mathbf{f}}}{\bar {\mathbf{f}}^{\sf H} \hat{\mathbf{B}}_{\mathrm{c}}(k) \bar {\mathbf{f}}}}\right) }\right)^{-\frac {1}{\alpha }} }\right \}^{-\frac {\alpha }{\log _{2} e}} \notag \\
&\qquad =\frac {\lambda _{\sf num} (\bar {\mathbf{f}})}{\lambda _{\sf den} (\bar {\mathbf{f}})}. \label{lambda}
\end{align}
\end{theorem}
\begin{proof}
    See Appendix C.
\end{proof}
We observe that \eqref{KKT condition} is interpreted as a form of non-linear eigenvector-dependent eigenvalue problem. The matrix ${\mathbf{B}}_{\sf KKT}^{-1} (\bar {\mathbf{f}}^{}){\mathbf{A}}_{\sf KKT}(\bar {\mathbf{f}}^{})$ is a nonlinear matrix function of $\bar {\mathbf{f}}$, and its eigenvalue $\lambda (\bar {\mathbf{f}})$ is set as the objective function of our main problem \eqref{mainProblem}.
Crucially, if we find the leading eigenvector, denoted as $\bar {\bf{f}}^{\star }$, it naturally maximizes our objective function of \eqref{mainProblem} while satisfying the condition \eqref{KKT condition}.


To obtain $\bar {\bf{f}}^{\star}$, we propose a GPI-based algorithm. 
To be specific, denoting that $\bar {\bf{f}}_{(t)}$ as a precoding vector obtained in $t$-th iteration, 
we iteratively update the precoding vector by 
\begin{align} \bar {\mathbf{f}}_{(t)} \leftarrow \frac {{\mathbf{B}}_{\sf KKT}^{-1} (\bar {\mathbf{f}}_{(t-1)}) {\mathbf{A}}_{\sf KKT} (\bar {\mathbf{f}}_{(t-1)}) \bar {\mathbf{f}}_{(t-1)}}{\| {\mathbf{B}}_{\sf KKT}^{-1} (\bar {\mathbf{f}}_{(t-1)}) {\mathbf{A}}_{\sf KKT} (\bar {\mathbf{f}}_{(t-1)}) \bar {\mathbf{f}}_{(t-1)} \|},\end{align}
where ${\mathbf{A}}_{\sf KKT}(\bar {\mathbf{f}}_{(t-1)}) $, ${\mathbf{B}}_{\sf KKT}(\bar {\mathbf{f}}_{(t-1)}) $ are calculated as \eqref{A_KKT} and \eqref{B_KKT}.
Note that this process is repeated until the update of precoder is bounded by predefined parameter, which is $\left \|{\bar {\mathbf{f}}_{(t)} - \bar {\mathbf{f}}_{(t-1)} }\right \| < \epsilon$. 

{\color{black}{
\begin{remark}\normalfont(Rationales for RSMA and extension to NOMA)  
While we mainly consider RSMA, non-orthogonal multiple access (NOMA) is another multiple access technique robust with MUI. 
Nonetheless, 
in MIMO-NOMA, it is difficult to find the optimal SIC order because each user's SE is determined not only by its channel strength, but also by a complicated function of channel direction, transmit power, and precoding vectors \cite{RSMA-ten-promising}. 
For this reason, as shown in \cite{clerks:critical:21}, MIMO-NOMA achieves mediocre SE, even lower than that of conventional SDMA. 
In contrast, RSMA achieves high SE, particularly in imperfect CSIT scenarios. This is thanks to its unique message construction, by which SIC ordering is simplified \cite{RSMA-ten-promising}. 
This is a primary reason that we choose RSMA instead of NOMA to mitigate MUI. 
However, we note that our approach also can be extended to MIMO-NOMA, which presents interesting future work. 
\end{remark}}}

\section{Simulation Results} \label{sec:sim}
In this section, we numerically examine the actual MSE of 2D-NOMP in DL channel reconstruction, compare it with the expected MSE from our ECM estimation. We also present the ergodic sum SE performance comparison between the proposed method and other baseline methods in various environments. 

\subsection{MSE of DL Channel Reconstruction}
\begin{figure}
\centering
\includegraphics[width=1\columnwidth]{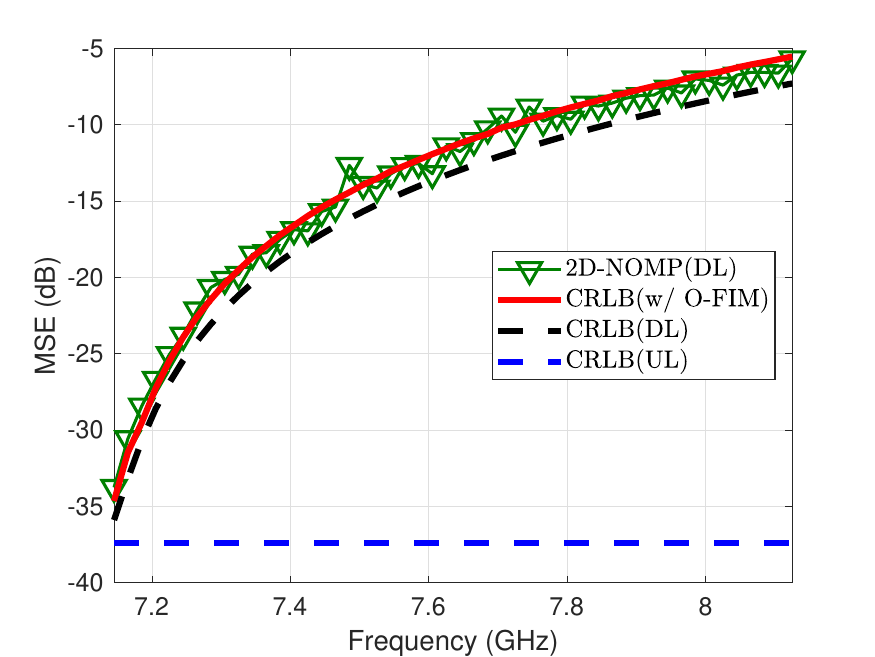}
\caption{MSE over DL carrier frequency. The UL carrier frequency is 7.15 GHz and UL SNR is $10$ dB with $M = 128$.}\label{expolRange}
\end{figure}

We first define the MSE of estimated DL channel of user $k$, i.e. $\hat{\bf{h}}_k(f)$ as  
\begin{align} \label{eq:mse}
\mathrm{MSE} = \mathbb{E}\left[{\|\mathbf{h}_k(f)-\hat{\mathbf{h}}_k(f)\|^2}\right].
\end{align}
Based on this, we plot the actual MSE \eqref{eq:mse} of the 2D-NOMP algorithm as a function of the frequency difference $f$ in Fig. \ref{expolRange}. 
We then compare the analytical MSE obtained by the CRLB and by our ECM estimation in \eqref{approx-w/o-disc}.
To yield the MSE from the CRLB and estimated ECM, we compute $\mathrm{tr}\{\mathbf{C}(f)\}$ and $\mathrm{tr}\{\hat {\boldsymbol{\Phi}}_k\}$ by following \cite{Rottenberg:2020}. 

In Fig. \ref{expolRange}, we first observe that the actual MSE of 2D-NOMP is close to the level of CRLB, i.e., $\mathrm{tr}\{\mathbf{C}(f)\}$.
This indicates that 2D-NOMP achieves near-optimal MSE performance in DL channel reconstruction, which is consistent with findings in \cite{NOMP}. 
Notably, we find that the expected MSE from our proposed ECM estimation $\mathrm{tr}\{\hat {\boldsymbol{\Phi}}_k\}$ is closer to the actual MSE of the 2D NOMP estimator than the original CRLB case. This result justifies the use of O-FIM for ECM estimation, not only because of its practical applicability but also its accuracy.

\subsection{Ergodic Sum SE}
\begin{figure}[t]
 \begin{subfigure}
 \centering
 \includegraphics[width=1\columnwidth]{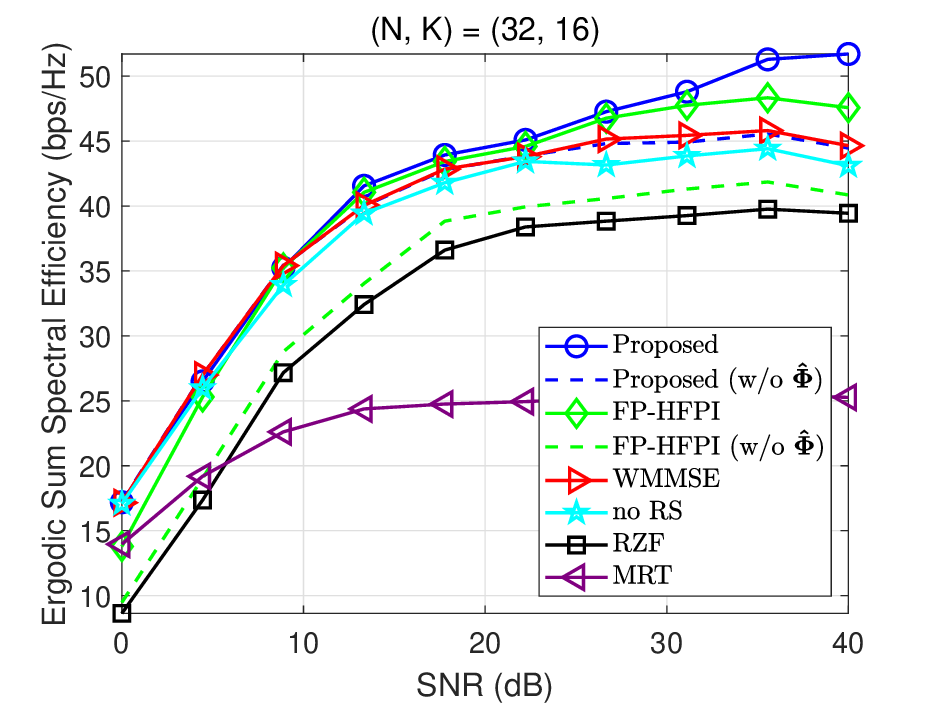}    
  \caption*{(a) \textcolor{black}{Reciprocal channel gain ($\eta_{k, \ell}^2=1, \forall k, \ell$).}}
 \label{fig:h1}
  \end{subfigure}
  \begin{subfigure}
  \centering 
  \includegraphics[width=1\columnwidth]{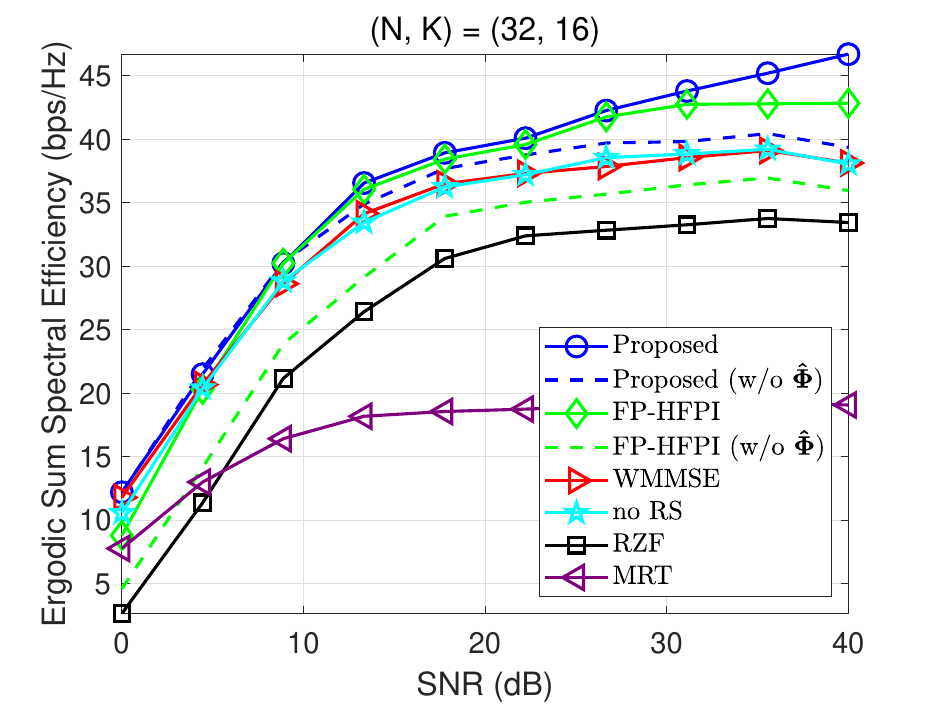} 
 \caption*{(b) \textcolor{black}{Non-reciprocal channel gain ($\eta_{k,\ell}^2\sim \mathrm{Unif}(\mu,1), \forall{k, \ell}$).}}
  \label{fig:h2}
  \end{subfigure}
\caption{\textcolor{black}{Comparison of the ergodic sum SE over transmit SNR, i.e., $P/\sigma^2$ with $(N, K) = (32, 16)$ and $M=128$. The UL carrier frequency is 7.25 GHz, and DL carrier frequency is 7.75 GHz. We use $\epsilon = 0.1$, $\alpha = 0.1$ and $\mu = 0.9$. For the cases without $\hat{\bf{\Phi}}$, we assume $\hat{\bf{\Phi}} = {\bf{0}}$.}}
\label{mainFigure1}
\end{figure}
We now investigate the ergodic sum SE of the proposed method through numerical simulations. 
For the simulation environments, we follow the system model laid forth in Section \ref{sec:model}. As baseline methods, we consider the following. We assume that the reconstructed DL channel \eqref{estDlChannel} is used for each baseline precoding method. 
\begin{itemize}
    \item \textcolor{black}{\textbf{FP-HFPI}: As a state-of-the-art RSMA precoding method, we consider the FP-HFPI method, wherein the precoder optimization is performed based on FP \cite{mao:twc:24}. For a comprehensive comparison, we incorporate our ECM estimation $\hat {\bf{\Phi}}_k$ into the FP-HFPI method.}
    \item \textbf{WMMSE}: As a state-of-the-art SDMA precoding optimization method, we consider WMMSE \cite{wmmse:twc:08}.
    \item {\textbf{no RS}}: 
    We consider the precoding method that maximizes the sum SE by adopting classical SDMA \cite{Jiwook}. 
    \item \textbf{MRT}: The precoders are aligned by the estimated channel vector, i.e. $\mathbf{f}_k=\hat{\mathbf{h}}_k, k\in\CMcal{K}$.
    \item \textbf{RZF}: The precoders are designed in a regularized zero-forcing (RZF) fashion as
    \begin{align}
        \mathbf{f}_k= \left( \hat{\mathbf{H}}\hat{\mathbf{H}}^{\sf H}+\frac{\sigma^2}{P}\mathbf{I} \right)^{-1}\hat{\mathbf{h}}^{\sf H}_k,
    \end{align}
    where $\hat{\mathbf{H}}$ denotes the matrix formed by stacking all the users' estimated channels \cite{heath-lozano}.
\end{itemize}

In Fig. \ref{mainFigure1}, we show the ergodic sum SE for the proposed method and the baseline methods, assuming \textcolor{black}{$\eta_{k, \ell}^2=1, \forall{k, \ell}$} for Fig.~\ref{mainFigure1}-(a) and \textcolor{black}{$\eta^2_{k,\ell} \sim \mathrm{Unif(\mu,1)}, \forall{k,\ell}$} for Fig.~\ref{mainFigure1}-(b). First, we observe that the RSMA approaches offer significant SE gains over SDMA approaches, especially in mid-to-high SNR regimes. \textcolor{black}{For instance, with reciprocal channel gain, the proposed method improves SE by up to 15.8\% over the SDMA method in \cite{Jiwook}. In the non-reciprocal channel gain case, which causes inaccurate CSIT and typically reduces SE performance, the performance gap between our method and \cite{Jiwook} increases up to 22.7\% at SNR of 40 dB, demonstrating the effective MUI mitigation capability of RSMA.}

{\color{black}{Compared to FP-HFPI, our method achieves better sum SE performance at low and high SNR regimes. Especially, at SNR of 40 dB in Fig. 2-(b), the proposed method attains around 9\% improvement in sum SE over FP-HFPI.
We interpret these gains as stemming from the proposed method’s ability to capture the impacts of the ECM. That is to say, our method exploits the SINR expression in its original form, while FP-HFPI alters it via a quadratic transform \cite{mao:twc:24}. This may allow the proposed method to more suitably reflect the impacts of the ECM into the precoder design. 
\textcolor{black}{It is crucial to note that, this level of SE performance for FP-HFPI is only achievable when our ECM estimation is properly incorporated. If FP-HFPI does not rely on our ECM estimation, Fig. 2-(b) shows that the performance gap between the proposed method and FP-HFPI increases to $23.2\%$ at SNR of 40dB. At other SNR levels in Fig. 2-(a), (b), FP-HFPI suffers from significant SE degradation when the ECM estimation is not used. 
}}}

These observations indicate that the ECM is crucial for achieving robust SE.
\textcolor{black}{For example, in Fig.~\ref{mainFigure1}-(b), incorporating the ECM into the proposed precoder design results in 21.1\% improvement in terms of sum SE at the SNR of 40 dB.}
\textcolor{black}{This highlights the need for careful handling of CSIT errors to fully realize RSMA’s potential. }

\textcolor{black}{\subsection{Robustness in Various Channel Environments}}

\begin{table}[!t]
  \begin{center}
    \caption{Relative performance compared with perfect CSIT}
    \label{tab:table1}
    \renewcommand{\arraystretch}{1.3} 
    \begin{tabular}{c|c|c|c|c} 
    \hline
    \multicolumn{5}{c}{(a) $(N, K)$ = (64, 32)} \\ \hline
    \text{Number of paths} & 1 & 4 & 7 & 10  \\ \hline
    \textbf{Proposed} (\%)& 96.01 & 93.00 & 89.00 & 83.63 \\ \hline
    \textbf{no RS} (\%)& 95.12 & 85.50 & 76.12 & 68.84 \\ \hline \hline
    \multicolumn{5}{c}{(b) $(N, K)$ = (16, 8)} \\ \hline
    \text{Number of paths} & 1 & 4 & 7 & 10  \\ \hline
    \textbf{Proposed} (\%)& 86.95 & 79.63 & 76.01 & 73.73 \\ \hline
    \textbf{no RS} (\%)& 86.51 & 65.36 & 58.34 & 50.98 \\ \hline
    \end{tabular}\label{tab:path}
  \end{center}
  \normalfont{Percentage of the sum SE performance compared to WMMSE under the assumption of perfect CSIT \cite{wmmse:twc:08}, for different numbers of channel paths at an SNR of 20 dB.} 
\end{table}

\begin{figure} [t]
\centering
\includegraphics[width=0.9\columnwidth]{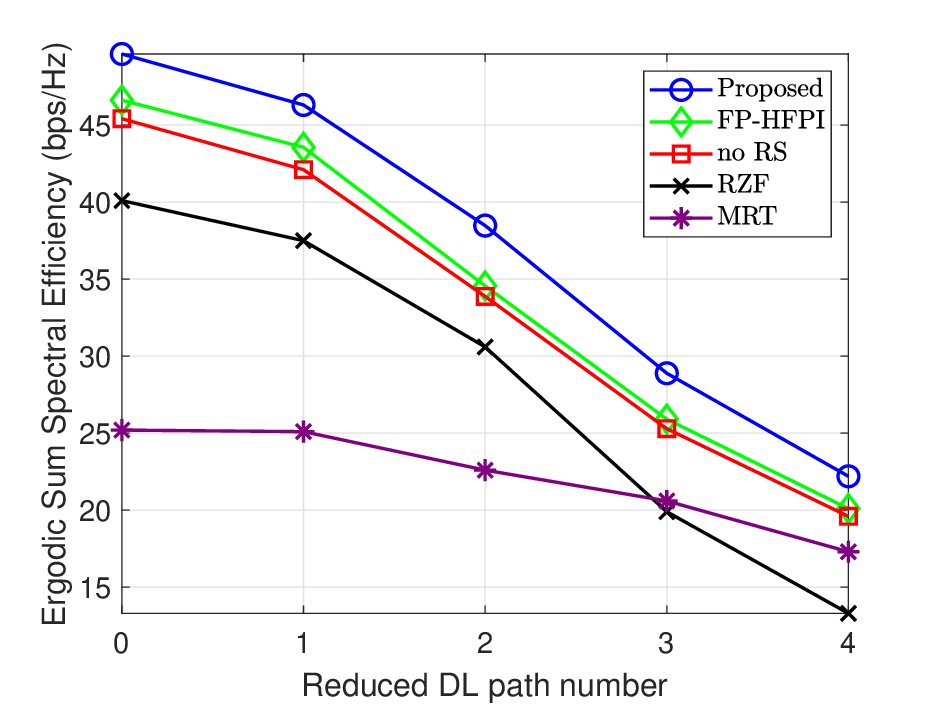}
\caption{\color{black}{Comparison of the ergodic sum SE over reduced DL path number with SNR = 30 dB. It is assumed that the UL channel consists of 5 paths and other simulation settings are identical to those of Fig.~\ref{mainFigure1}-(b).} 
}\label{mainFigure3}
\end{figure}

We explore the robustness of the proposed method in various settings.
In Table I, we compare the sum SE achieved by our approach versus WMMSE with perfect CSIT \cite{wmmse:twc:08}. 
Since our approach does not rely on any CSI feedback, Table I implies the inevitable performance loss resulting from its absence. 
In general, the performance gap decreases as $(N,K)$ increases and the number of paths decreases.
{\color{black}{This is reasonable because, as $N$ increases, it enables the accurate parameter estimation during UL training, which is effective in mitigating MUI.
Similarly, as the number of channel paths decreases, the channels from different users can become nearly orthogonal, which reduces the MUI and thereby increases the sum SE. 
Given that the number of dominant DL paths is typically small ($\sim$4) as demonstrated in \cite{drctnl:twc:18}, and also perfect CSIT is not feasible even by using CSI feedback, 
the actual performance gap in practice would be much less than $7$\% for $(N,K) = (64, 32)$.}}

\begin{figure*}[t]
\centering
\includegraphics[width=0.7\textwidth]{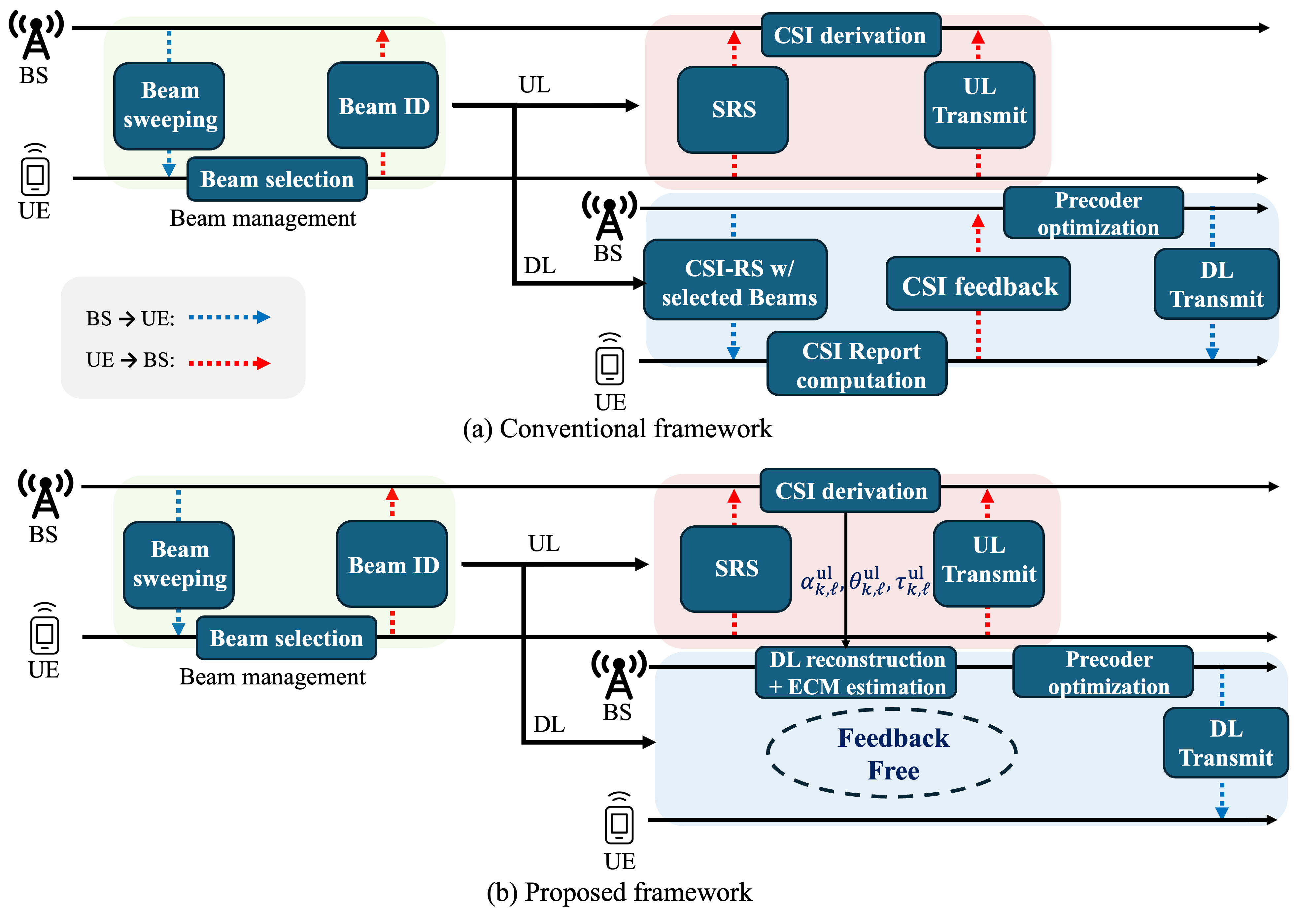}
\caption{Comparison of conventional FDD MU-MIMO DL CSI acquisition process vs. the proposed CSI acquisition process.}\label{CE_process}
\end{figure*}
{\color{black}{ 
Through the comparison, we provide a trade-off guideline between the latency reduction from saving CSI feedback and the corresponding loss of sum SE. In conventional DL transmission within FDD systems, the users estimate the DL channel using reference signals (e.g., CSI-RS) and send feedback (e.g., PMI, RI, SINR) to the BS. 
In this process, the primary latency arises from CSI report computation on the user side, which can take up to 6 ms due to limited computational resources (\cite{3GPP}, Table 5.4-2). This is much longer than other steps, such as CSI-RS scheduling ($\sim$1 ms), CSI feedback ($\sim$1 ms), and decoding ($\sim$1 ms). 
Using our approach, we shift this CSI processing overhead from the user side to the BS (see Fig. \ref{CE_process}). This can save approximately 6-7 ms, while achieving more than $93$\% of the sum SE in $(N, K) = (64, 32)$ \textcolor{black}{assuming 4 dominant paths \cite{drctnl:twc:18}}. As mentioned before, since perfect CSIT cannot be guaranteed even with conventional CSI feedback \cite{guo:tcom:22}, the de facto performance loss of the proposed approach in sum SE is well below $7$\%. As a result, our approach offers a low-latency MIMO transmission framework that significantly reduces latency with only a slight impact on sum SE. 
We illustrate the comparison of CSI acquisition processes in Fig.~\ref{CE_process}. 


} }

{\textcolor{black}{Fig.~\ref{mainFigure3} shows the sum SE performance when the number of DL channel path reduces, which occurs due to a higher carrier frequency of the DL band in FDD \cite{FDD-measurement}.
This discrepancy between actual channel path numbers between UL and DL naturally leads to performance degradation, as it would result in inaccurate channel reconstruction. For all the cases of reduced DL path, the proposed method provides robust SE performance compared to other SDMA methods. Similar to the above observation, this robustness of our method is attributed to the proper ECM estimation. }}

\section{Conclusion}
In this paper, we have proposed a novel method to achieve robust SE performance in FDD massive MIMO systems without the need for CSI feedback. 
In order to effectively manage MUI caused by imperfect DL channel reconstruction, our method uses RSMA. This approach requires carefully integrating the CSIT error into the precoder design, with the proposed ECM estimation being crucial.
Our major findings are as follows. 
i) The proposed ECM estimation using the O-FIM accurately captures DL CSIT reconstruction errors. 
ii) The proposed method significantly improves the sum SE performance over the existing methods. In particular, accurate ECM estimation is crucial for obtaining these gains. 
iii) Our approach offers a low-latency FDD massive MIMO transmission framework, achieving a favorable trade-off between significant latency reduction and a slight loss in sum SE.
As future work, it is promising to extend our approach by considering various system environments, such as energy efficiency maximization, max-min fairness, and integrated sensing and communications. 

\section*{Appendix A}
\section*{Proof of Lemma 1}
Let us assume each measurement $\mathbf{y}_{n, m}$ at each antenna and sub-carrier ($n \in [N], m \in [M]$) is drawn under independent and identically distributed with likelihood $f(\mathbf{y}_{n,m}|\boldsymbol{\psi}_k)$, and its distribution follows Gaussian distribution with covariance $\sigma^2\mathbf{I}$ and mean $\bar{\mathbf{y}}$.
Then, the log-likelihood of the parameters $\boldsymbol{\psi}_k$ given measured signal can be represented by
\begin{align}
    \CMcal{L}(\boldsymbol{\psi}_k|\mathbf{y}_{n, m}, n \in [N], m \in [M]) &\triangleq \sum_{n=1}^{N} \sum_{m=1}^{M} \log f(\mathbf{y}_{n, m}|\boldsymbol{\psi}_k) \\
    &= -\frac{1}{\sigma^2}(\mathbf{y} - \bar{\mathbf{y}} )^{\sf{H}}(\mathbf{y} - \bar{\mathbf{y}}) + C,
\end{align} 
where $C$ is a constant. Then, the observed information defined as the negative Hessian of log-likelihood is evaluated at channel parameters $\hat{\boldsymbol{\psi}}_k$, as follows \cite{observedFisher}: 
\begin{align}
    & \boldsymbol{\mathbf{I}}(\hat{\boldsymbol{\psi}}_k) = -\nabla\nabla^{\sf{H}}\CMcal{L}(\boldsymbol{\psi}_k)|_{\boldsymbol{\psi}_k = \hat{\boldsymbol{\psi}}_k} \\
    & = -\left.\begin{bmatrix} \label{Observed info}
    \frac{\partial^2}{\partial^2\psi_{1,1}} & \frac{\partial^2}{\partial \psi_{1,1} \partial \psi_{1,2}} & \cdots & \frac{\partial^2}{\partial \psi_{1,1} \partial \psi_{L, 4}} \\
    \frac{\partial^2}{\partial \psi_{1,2}\partial \psi_{1,1}} & \frac{\partial^2}{\partial \psi_{1,2}^2} & \cdots & \frac{\partial^2}{\partial \psi_{1,2} \partial \psi_{L, 4}} \\
    \vdots & \vdots & \ddots & \vdots \\
    \frac{\partial^2}{\partial \psi_{L, 4} \partial \psi_{1, 1}} & \frac{\partial^2}{\partial \psi_{L,4} \partial \psi_{1,2}} & \cdots & \frac{\partial^2}{\partial^2 \psi_{L, 4}} \\
\end{bmatrix} \CMcal{L}(\boldsymbol{\psi}_k) \right \rvert_{\boldsymbol{\psi_k} = \hat{\boldsymbol{\psi}}_k},
\end{align}
where $\CMcal{L}(\boldsymbol{\psi}_k)$ denotes $\CMcal{L}(\boldsymbol{\psi}_k|\mathbf{y}_{n, m}, n \in [N], m \in [M])$ for simplicity and $\CMcal{L}(\boldsymbol{\psi}_k)|_{\boldsymbol{\psi_k} = \hat{\boldsymbol{\psi}}_k}$ follows the normal distribution as well. 
To evaluate $(u, v)$ component of $\boldsymbol{\mathbf{I}}(\hat{\boldsymbol{\psi}}_k)$, 
we compute
\begin{align*}
    \frac{\partial\CMcal{L}(\boldsymbol{\psi}_k)}{\partial\psi_v} = \frac{1}{\sigma^2}\sum_{n=1}^{N}\sum_{m=1}^{M}\left(\frac{\partial\bar{\mathbf{y}}_{n,m}}{\partial\psi_v}\right)^{*}(\mathbf{y}_{n,m}-\bar{\mathbf{y}}_{n,m}) \\
    +\frac{1}{\sigma^2}\sum_{n=1}^{N}\sum_{m=1}^{M}(\mathbf{y}_{n,m}-\bar{\mathbf{y}}_{n,m})^{*}\left(\frac{\partial\bar{\mathbf{y}}_{n,m}}{\partial\psi_v}\right), \\
    = \frac{2}{\sigma^2}\textrm{Re} \left\{\sum_{n=1}^{N}\sum_{m=1}^{M}(\mathbf{y}_{n,m}-\bar{\mathbf{y}}_{n,m})^{*}\left(\frac{\partial\bar{\mathbf{y}}_{n,m}}{\partial\psi_v}\right)\right\},
\end{align*}
where $\bar{\mathbf{y}}_{n,m}$ denotes the reconstructed signal at $n$-th antenna and $s$-th sub-carrier. Then, the derivative with respect to $\psi_u$ can be found as follows
\begin{align} \label{Hessian}
 \frac{\partial}{\partial\psi_u}\left(\frac{\partial\CMcal{L}(\boldsymbol{\psi}_k)}{\partial\psi_v}\right) = \frac{2}{\sigma^2} &\textrm{Re} 
 \left\{\sum_{n=1}^{N}\sum_{m=1}^{M}\left(-\frac{\partial\bar{\mathbf{y}}^{*}_{n,m}}{\partial\psi_u}\frac{\partial\bar{\mathbf{y}}_{n,m}}{\partial\psi_v}\right) \right. \\
 \quad &+\left.\sum_{n=1}^{N}\sum_{m=1}^{M} (\mathbf{y}_{n,m}-\bar{\mathbf{y}}_{n,m})^{*}\frac{\partial^2\bar{\mathbf{y}}_{n,m}}{\partial\psi_u\partial\psi_v}
 \right\}.
\end{align}
Substituting \eqref{Hessian} into \eqref{Observed info} completes the proof. 

\section*{Appendix B}
\section*{Proof of Corollary 1}
\textcolor{black}{Consider first the basic case of $\alpha_{k, \ell}^{\mathrm{dl}}=\alpha_{k, \ell}^{\mathrm{ul}}, \forall k, \ell$, then we would represent the ECM of reconstructed channel in \eqref{dlChannel} as
\begin{align}
\mathbf{\Phi}_k \triangleq \mathbb{E}_{\{\mathbf{e}_k|\hat{\boldsymbol{\psi}}_k\}}\Bigg[
    &\left(\sum_{\ell=1}^{L_k}\alpha_{k,\ell}^{\mathrm{dl}}\bar{\mathbf{u}}_{k,\ell}
    - \sum_{\ell=1}^{L_k}\hat{\alpha}_{k,\ell}^{\mathrm{dl}}\hat{\bar{\mathbf{u}}}_{k,\ell}\right) \notag \\
    &\times \left(\sum_{\ell=1}^{L_k}\alpha_{k,\ell}^{\mathrm{dl}}\bar{\mathbf{u}}_{k,\ell}
    - \sum_{\ell=1}^{L_k}\hat{\alpha}_{k,\ell}^{\mathrm{dl}}\hat{\mathbf{u}}_{k,\ell}\right)^{\sf H}
\Bigg],
\end{align}
where we define $\bar{\mathbf{u}}_{k, \ell}$ as
\begin{align}
    \bar{\mathbf{u}}_{k, \ell}=\mathbf{a}\left(\theta^{\mathrm{dl}}_{k,\ell};\lambda^{\mathrm{dl}}\right)e^{-j2\pi f\tau_{k,\ell}} \in \mathbb{C}^{N\times 1},
\end{align}which represents the sampled signature of $\mathbf{u}_{k, \ell}$ from \eqref{u_def} at frequency $f$, and let $\hat{\bar{\mathbf{u}}}$ denote its estimate. Recall that this can be obtainable via \eqref{approx-w/o-disc}.
Based on this, it can be extended to the general case: $\alpha^{\mathrm{dl}}_{k, \ell}=\eta_{k, \ell}\alpha^{\mathrm{ul}}_{k, \ell}+\sqrt{1-\eta^2_{k, \ell}}\beta_{k, \ell}$, where $\beta_{k, \ell} \sim \mathcal{CN}(0,\sigma^2_{\mathrm{path}, k})$ and independent of $\alpha_{k, \ell}^{\mathrm{ul}}, \hat{\alpha}_{k, \ell}^{\mathrm{ul}}$. Then, we possibly develop the new ECM as follows:
\begin{align}
\mathbb{E} \Bigg[
    \left(\sum_{\ell=1}^{L_k}(\eta_{k,\ell}\alpha_{k,\ell}^{\mathrm{ul}}+\sqrt{1-\eta^2_{k, \ell}}\beta_{k, \ell})\bar{\mathbf{u}}_{k,\ell}
    - \sum_{\ell=1}^{L_k}\eta_{k,\ell}\hat{\alpha}_{k,\ell}^{\mathrm{ul}}\hat{\bar{\mathbf{u}}}_{k,\ell}\right) \notag \\
    \times \left(\sum_{\ell=1}^{L_k}(\eta_{k,\ell}\alpha_{k,\ell}^{\mathrm{ul}}+\sqrt{1-\eta^2_{k, \ell}}\beta_{k, \ell})\bar{\mathbf{u}}_{k,\ell}
    - \sum_{\ell=1}^{L_k}\eta_{k,\ell}\hat{\alpha}_{k,\ell}^{\mathrm{ul}}\hat{\bar{\mathbf{u}}}_{k,\ell}\right)^{\sf H}
\Bigg],\label{newECM}
\end{align}
where we assumed that $\hat{\alpha}^{\mathrm{dl}} = \eta_{k,\ell}\hat{\alpha}^{\mathrm{ul}}_{k,\ell}$ using the correlation knowledge. 
To gain further insight, we can consider the following assumptions by examining \eqref{newECM}, i.e.,
\begin{align}
&\mathbb{E}\left[\left(\alpha_{k,\ell}^{\mathrm{ul}}\bar{\mathbf{u}}_{k,\ell}
-\hat{\alpha}_{k,\ell}^{\mathrm{ul}}\hat{\bar{\mathbf{u}}}_{k,\ell}\right)
\left(\alpha_{k',\ell'}^{\mathrm{ul}}\bar{\mathbf{u}}_{k',\ell'}
-\hat{\alpha}_{k',\ell'}^{\mathrm{ul}}\hat{\bar{\mathbf{u}}}_{k',\ell'}\right)^{\sf H}\right] \approx 0,\nonumber \\
&\qquad \qquad \qquad \qquad \qquad \qquad \qquad \forall k\neq k' \text{ or } \ell \neq \ell',
\end{align}
which implies no error correlation between different user or path.
In addition, by leveraging the relationship between $\beta_{k,\ell}$ and $\alpha_{k,\ell}, \hat{\alpha}_{k,\ell}$, \eqref{newECM} can be developed by
\begin{equation}
    \begin{aligned}
    &\mathbb{E}\Bigg[\sum_{\ell=1}^{L_k}\eta^2_{k,\ell}(\alpha_{k,\ell}^{\mathrm{ul}}\bar{\mathbf{u}}_{k,\ell}-\hat{\alpha}_{k,\ell}^{\mathrm{ul}}\hat{\bar{\mathbf{u}}}_{k,\ell})(\alpha_{k,\ell}^{\mathrm{ul}}\bar{\mathbf{u}}_{k,\ell}-\hat{\alpha}_{k,\ell}^{\mathrm{ul}}\hat{\bar{\mathbf{u}}}_{k,\ell})^{\sf H} \\
    &\qquad \qquad \quad \qquad \quad \quad + \sum_{\ell=1}^{L_k}(1-\eta_{k,\ell}^2)\beta_{k,\ell}\beta_{k, \ell}^{\sf H}\bar{\mathbf{u}}_{k,\ell}\bar{\mathbf{u}}_{k,\ell}^{\sf H}\Bigg] \\
    &= \left(\sum_{\ell=1}^{L_k}\frac{\eta^2_{k,\ell}}{L_k}\right)\mathbf{\Phi}_k+\left(\sum_{\ell=1}^{L_k}\frac{1-\eta_{k,\ell}^2}{L_k}\right)\mathbf{I}_{n}\label{eq:trueECM},
    \end{aligned}
\end{equation}
where $\sigma_{\mathrm{path}, k}^2 = \frac{1}{NL_k}$ by normalization. Since the underlying ECM given channel model follows \eqref{eq:trueECM}, we replace $\boldsymbol{\Phi}_k$ with $\tilde{\mathbf{C}}(f)$ to estimate the ECM. This completes the proof.}
\section*{Appendix C}
\section*{Proof of Theorem 1}
We first explore the KKT condition of our main problem (\ref{mainProblem}). In this, we develop the Lagrangian function as
\begin{align} L(\bar{\mathbf{f}}) = \log \left( \frac{1}{K} \sum_{k \in [K]} \exp \left( \log_2 \left( \frac{\bar{\mathbf{f}}^{\sf H} \mathbf{A}_{\mathrm{c}}(k) \bar{\mathbf{f}}}{\bar{\mathbf{f}}^{\sf H} \mathbf{B}_{\mathrm{c}}(k) \bar{\mathbf{f}}} \right) \right)^{-\frac{1}{\alpha}} \right)^{-\alpha} \notag \\
+ \sum_{k = 1}^{K} \log_2 \left( \frac{\bar{\mathbf{f}}^{\sf H} \mathbf{A}_k \bar{\mathbf{f}}}{\bar{\mathbf{f}}^{\sf H} \mathbf{B}_k \bar{\mathbf{f}}} \right). \label{Lagrangian}\end{align}
Then, we take the partial derivatives of $L(\bar{\mathbf{f}})$ in terms of $\bar{\mathbf{f}}$ to find a stationary point, where we set it as zero.
For notational simplicity, we denote the first term of (\ref{Lagrangian}) as $L_{1}(\bar{\mathbf{f}})$ and second term as $L_{2}(\bar{\mathbf{f}})$ respectively, and each partial derivative is obtained respectively, as follows
\begin{align} 
&\frac {\partial L_{1}(\bar {\mathbf{f}})}{\partial \bar {\mathbf{f}}^{\sf H}}={ { \sum _{k \in [K]} \left ({\frac {\exp \left ({\frac {1}{-\alpha } \frac {\bar {\mathbf{f}}^{\sf H} {\mathbf{A}}_{\mathrm{c}}(k) \bar {\mathbf{f}}}{\bar {\mathbf{f}}^{\sf H} {\mathbf{B}}_{\mathrm{c}}(k) \bar {\mathbf{f}}} }\right)}{\sum _{j \in [K]} \exp \left ({\frac {1}{-\alpha } \log _{2} \Bigg ({\frac {\bar {\mathbf{f}}^{\sf H} {\mathbf{A}}_{\mathrm{c}}(j) \bar {\mathbf{f}}}{\bar {\mathbf{f}}^{\sf H} {\mathbf{B}}_{\mathrm{c}}(j) \bar {\mathbf{f}}} }\Bigg) }\right) } }\right) }} \nonumber \\
&\qquad \quad \quad \quad \times \, \partial \left ({\log _{2}\left ({\frac {\bar {\mathbf{f}}^{\sf H} {\mathbf{A}}_{\mathrm{c}}(k) \bar {\mathbf{f}}}{\bar {\mathbf{f}}^{\sf H} {\mathbf{B}}_{\mathrm{c}}(k) \bar {\mathbf{f}}}}\right) }\right)\!/\!\partial {\mathbf{f}}^{\sf H} 
\end{align}
\begin{align}
&=\frac {1}{\log 2} { \sum _{i = 1}^{G} \left \{{ \sum _{k \in [K]} \left ({\frac {\exp \left ({\frac {1}{-\alpha } \frac {\bar {\mathbf{f}}^{\sf H} {\mathbf{A}}_{\mathrm{c}}(k) \bar {\mathbf{f}}}{\bar {\mathbf{f}}^{\sf H} {\mathbf{B}}_{\mathrm{c}}(k) \bar {\mathbf{f}}} }\right)}{\sum _{j \in [K]} \exp \left ({\frac {1}{-\alpha } \log _{2} \Bigg ({\frac {\bar {\mathbf{f}}^{\sf H} {\mathbf{A}}_{\mathrm{c}}(j) \bar {\mathbf{f}}}{\bar {\mathbf{f}}^{\sf H} {\mathbf{B}}_{\mathrm{c}}(j) \bar {\mathbf{f}}} }\Bigg) }\right) } }\right) }\right \}} \nonumber \\
& \qquad ~\quad \times \, \left \{{\frac {{\mathbf{A}}_{\mathrm{c}}(k) \bar {\mathbf{f}}}{\bar {\mathbf{f}}^{\sf H} {\mathbf{A}}_{\mathrm{c}}(k) \bar {\mathbf{f}}} - \frac {{\mathbf{B}}_{\mathrm{c}}(k) \bar {\mathbf{f}}}{\bar {\mathbf{f}}^{\sf H} {\mathbf{B}}_{\mathrm{c}}(k) \bar {\mathbf{f}}} }\right \}\Bigg],
\end{align}
\begin{align}
        \frac {\partial L_{2} (\bar {\mathbf{f}})}{\partial \bar {\mathbf{f}}^{\sf H}} =\frac {1}{\log 2} \sum _{k = 1}^{K} \left [{\frac {{\mathbf{A}}_{k} \bar {\mathbf{f}}}{\bar {\mathbf{f}}^{\sf H} {\mathbf{A}}_{k} \bar {\mathbf{f}}} - \frac {{\mathbf{B}}_{k} \bar {\mathbf{f}}}{\bar {\mathbf{f}}^{\sf H} {\mathbf{B}}_{k} \bar {\mathbf{f}}} }\right].
\end{align}
It is obvious that when the sum of both terms is zero, stationarity for first-order KKT condition is met, which is
\begin{align}
&\frac {\partial L_{1}(\bar {\mathbf{f}})}{\partial \bar {\mathbf{f}}^{\sf H}} + \frac {\partial L_{2}(\bar {\mathbf{f}})}{\partial \bar {\mathbf{f}}^{\sf H}} = 0\\\Leftrightarrow&{{ \sum _{k \in [K]} \left ({\frac {\exp \left ({\frac {1}{-\alpha } \frac {\bar {\mathbf{f}}^{\sf H} {\mathbf{A}}_{\mathrm{c}}(k) \bar {\mathbf{f}}}{\bar {\mathbf{f}}^{\sf H} {\mathbf{B}}_{\mathrm{c}}(k) \bar {\mathbf{f}}} }\right)}{\sum _{j \in [K]} \exp \left ({\frac {1}{-\alpha } \log _{2} \Bigg ({\frac {\bar {\mathbf{f}}^{\sf H} {\mathbf{A}}_{\mathrm{c}}(j) \bar {\mathbf{f}}}{\bar {\mathbf{f}}^{\sf H} {\mathbf{B}}_{\mathrm{c}}(j) \bar {\mathbf{f}}} }\Bigg) }\right) } }\right) }} \\&\qquad \qquad ~\quad \times \, \left \{{\frac {{\mathbf{A}}_{\mathrm{c}}(k) \bar {\mathbf{f}}}{\bar {\mathbf{f}}^{\sf H} {\mathbf{A}}_{\mathrm{c}}(k) \bar {\mathbf{f}}} - \frac {{\mathbf{B}}_{\mathrm{c}}(k) \bar {\mathbf{f}}}{\bar {\mathbf{f}}^{\sf H} {\mathbf{B}}_{\mathrm{c}}(k) \bar {\mathbf{f}}} }\right \}\Bigg] \\&+ \sum _{k = 1}^{K} \left [{\frac {{\mathbf{A}}_{k} \bar {\mathbf{f}}}{\bar {\mathbf{f}}^{\sf H} {\mathbf{A}}_{k} \bar {\mathbf{f}}} - \frac {{\mathbf{B}}_{k} \bar {\mathbf{f}}}{\bar {\mathbf{f}}^{\sf H} {\mathbf{B}}_{k} \bar {\mathbf{f}}} }\right] = 0.
\end{align}
With equation (\ref{A_KKT}), (\ref{B_KKT}), (\ref{lambda}), we can arrange the first-order KKT condition as
\begin{align}
{\mathbf{A}}_{\sf KKT}(\bar {\mathbf{f}}) \bar {\mathbf{f}}=&\lambda (\bar {\mathbf{f}}) {\mathbf{B}}_{\sf KKT} (\bar {\mathbf{f}}) \bar {\mathbf{f}} \Leftrightarrow {\mathbf{B}}_{\sf KKT} (\bar {\mathbf{f}})^{-1}{\mathbf{A}}_{\sf KKT}(\bar {\mathbf{f}}) \bar {\mathbf{f}} \\=&\lambda (\bar {\mathbf{f}}) \bar {\mathbf{f}}.
\end{align}
This completes the proof.
\bibliographystyle{IEEEtran}
\bibliography{main}

\end{document}